\documentclass[useAMS,usenatbib]{mn2e}
\usepackage{lscape,graphicx}
\newcommand{\apg}{\:^{>}_{\sim}\:}
\newcommand{\apl}{\:^{<}_{\sim}\:}
\newcommand{\cmjj}{\mbox{${\rm cm^{-2}}$}}
\newcommand{\etal}{et al.}

\newcommand{\kms}{\mbox{km\ s${^{-1}}$}}

\begin{document}


\title[Spatially Resolved Halo Gas Kinematics at
$z=0.4-0.8$]{Spatially Resolved Velocity Maps of Halo Gas Around Two
  Intermediate-redshift Galaxies\thanks{Based on data gathered with
    the 6.5 m Magellan Telescopes located at Las Campanas Observatory,
    and the NASA/ESA Hubble Space Telescope operated by the Space
    Telescope Science Institute and the Association of Universities
    for Research in Astronomy, Inc., under NASA contract NAS 5-26555}}
\author[Chen et al.]{Hsiao-Wen Chen$^{1}$\thanks{E-mail:hchen@oddjob.uchicago.edu}, Jean-Ren\'e Gauthier$^2$, Keren Sharon$^3$, Sean D.\ Johnson$^{1}$, \newauthor Preethi Nair$^{4}$, Cameron J.\ Liang$^{1}$ \\
  \\
  $^{1}${Department of Astronomy \& Astrophysics, and Kavli Institute
    for Cosmological Physics, University of Chicago, Chicago, IL 60637, USA } \\
  $^{2}${Cahill Center for Astronomy and Astrophysics, California Institute of Technology, Pasadena, CA 91125, USA }\\
  $^{3}${Department of Astronomy, University of Michigan, 500 Church
    Street, Ann Arbor, MI 48109, USA}\\
  $^{4}${Space Telescope Science Institute, Baltimore, MD 21218, USA}}

\pagerange{\pageref{firstpage}--\pageref{lastpage}} \pubyear{2011}

\maketitle

\label{firstpage}

\begin{abstract}

  Absorption-line spectroscopy of multiply-lensed QSOs near a known
  foreground galaxy provides a unique opportunity to go beyond the
  traditional one-dimensional application of QSO probes and establish
  a crude three-dimensional (3D) map of halo gas around the galaxy
  that records the line-of-sight velocity field at different locations
  in the gaseous halo.  Two intermediate-redshift galaxies are
  targeted in the field around the quadruply-lensed QSO
  HE\,0435$-$1223 at redshift $z=1.689$, and absorption spectroscopy
  along each of the lensed QSOs is carried out in the vicinities of
  these galaxies.  One galaxy is a typical, star-forming $L_*$ galaxy
  at $z=0.4188$ and projected distance of $\rho=50$ kpc from the
  lensing galaxy.  The other is a super-$L_*$ barred spiral at
  $z=0.7818$ and $\rho=33$ kpc.  Combining known orientations of the
  quadruply-lensed QSO to the two foreground galaxies with the
  observed Mg\,II $\lambda\lambda$ 2796,2803 absorption profiles along
  individual QSO sightlines has for the first time led to spatially
  resolved kinematics of tenuous halo gas on scales of $5-10$ kpc at
  $z>0.2$.  A Mg\,II absorber is detected in every sightline observed
  through the halos of the two galaxies, and the recorded absorber
  strength is typical of what is seen in previous close QSO--galaxy
  pair studies.  While the multi-sightline study confirms the unity
  covering fraction of Mg\,II absorbing gas at $\rho < 50$ kpc from
  star-forming disks, the galaxies also present two contrasting
  examples of complex halo gas kinematics.
  Different models, including a rotating disk, collimated outflows,
  and gaseous streams from either accretion or tidal/ram-pressure
  stripping, are considered for comparisons with the absorption-line
  observations, and infalling streams/stripped gas of width $\apg 10$
  kpc are found to best describe the observed gas kinematics across
  multiple sightlines.  In addition, the observed velocity dispersion
  between different sightlines offers a crude estimate of turbulence
  in the Mg\,II absorbing halo gas.  The observations presented here
  demonstrate that multiple-QSO probes enable studies of spatially
  resolved gas kinematics around distant galaxies, which provide key
  insights into the physical nature of circumgalactic gas beyond the
  nearby universe.

\end{abstract}


\section{INTRODUCTION}

A key element in theoretical studies of galaxy formation and evolution
is an accurate characterization of gas infall and outflows around
star-forming regions, two competing processes that regulate star
formation over cosmic time.  Capturing these processes in observations
is therefore of great interest and importance in validating our view
of how galaxies grow.  Since the pioneering work of Boksenberg \&
Sargent (1978), Boksenberg \etal\ (1980), and Bergeron (1986),
absorption spectroscopy of distant QSOs has been utilized as an
effective means of probing tenuous gas around galaxies.  For every
projected galaxy and QSO pair, the background QSO serves as a single
pencil beam to explore the line-of-sight gas distribution through the
galactic halo at the projected distance where the QSO appears.

While QSO absorption spectroscopy offers unparalleled sensitivities
for uncovering low-density gas, a single QSO spectrum does not yield a
two-dimensional map of halo gas around individual galaxies like
conventional 21\,cm observations (e.g.\ Chynoweth \etal\ 2008).
Studies of halo gas around distant galaxies have therefore relied on a
statistical approach to characterize the spatial distribution of
tenuous gas in galactic halos (e.g.\ Lanzetta \& Bowen 1990) and to
estimate a mean value of gas covering fraction (e.g.\ Lanzetta \etal\
1995; Chen \etal\ 2010a; Tumlinson \etal\ 2011) over an ensemble of
intervening galaxies.  However, details regarding the spatial
variation of gas density and kinematics remain unknown for individual
halos.  Knowing the kinematics of halo gas revealed in absorption-line
surveys bears significantly on all effort to characterize gas infall
and outflows around star-forming galaxies using absorption
spectroscopy (e.g.\ Faucher-Gigu\`ere \& Kere$\check{\rm s}$ 2011).

Recent galaxy survey data have revealed the ubiquitous presence of
outflows in star-forming galaxies at $z>0.7$, through observations of
blue-shifted Mg\,II\,$\lambda\lambda$ 2796,2803 self-absorption
against the UV light from star-forming regions (e.g.\ Weiner \etal\
2009; Rubin \etal\ 2010; Bordoloi \etal\ 2013).  Although the distance
of the outflowing material is unknown in these observations (but see
Rubin \etal\ 2011 and Martin \etal\ 2012 for two cases that exhibit
outflowing gas in emission out to $\sim 10$ kpc), such finding has
triggered several follow-up studies that attribute the majority of
metal-line absorbers (such as Mg\,II, C\,IV, and O\,VI) uncovered along
random sightlines to those high-speed outflows revealed through
self-absorption of UV light (e.g.\ Steidel \etal\ 2010; Chelouche \&
Bowen 2010; Nestor \etal\ 2011; Tumlinson \etal\ 2011; Werk \etal\
2013).  Such interpretation naturally implies a minimal presence of
gas accretion around star-forming galaxies.

An additional empirical finding that supports the notion of a
non-negligible fraction of metal-line absorbers originating in
starburst driven outflows is the enhanced Mg\,II absorption near the
minor axes and within 50 projected kpc of disk galaxies at $z\sim 0.7$
by Bordoloi \etal\ (2011).  This finding has been followed by reports
of a possible bimodal azimuthal dependence of Mg\,II absorbers (e.g.\
Bouch\'e \etal\ 2012; Kacprzak \etal\ 2012), attributing metal-line
absorbers observed near minor axes to outflowing gas and those
observed near major axes to infalling gas.  A bimodal distribution in
the disk orientation of a Mg\,II-selected galaxy sample suggests that
both gas infall and outflows contribute comparably to the absorber
population.  It also suggests that the physical origin of an absorber
can be determined if the disk orientation is known.  However, the
velocity field of outflows/accretion is not known and such report has
also raised new questions.

For example, a natural expectation for absorption lines produced in
outflows is that the observed velocity profile depends on the
inclination of the star-forming disk, with the largest velocity spread
expected when looking directly into a face-on star-forming disk (e.g.\
Gauthier \& Chen 2012).  While such inclination-dependent absorption
width is clearly seen in the self-absorption of galaxy UV light (e.g.\
Kornei \etal\ 2012; Bordoloi \etal\ 2012), it appears to be weak or
absent among random absorbers found in transverse direction from
star-forming galaxies (e.g.\ Bordoloi \etal\ 2011; Bouch\'e \etal\
2012).  The lack of correlation between absorber width and disk
inclination appears to be discrepant from the expectations of an
outflow origin.  In addition, galactic-scale outflows in local
starbursts are observed to follow the path of least resistance along
the polar axis (e.g.\ Heckman \etal\ 1990).  If this feature also
applies to distant star-forming galaxies, then known disk orientation
and inclination allow us to deproject the observed line-of-sight
velocity distribution along the polar axis and examine the energetics
required to power the outflows.  Gauthier \& Chen (2012) showed that
if the Mg\,II absorbers observed at $\rho>7$ kpc from star-forming
galaxies originate in outflows, then either the outflows are
decelerating (inconsistent with the interpretation of blue-shifted
absorption tails by Martin \& Bouch\'e 2009 and Steidel \etal\ 2010)
or there needs to be additional kinetic energy input at $> 10$ kpc
beyond the disk plane.  Finally, Chen (2012) showed that both the
spatial extent and mean absorption equivalent width of halo gas around
galaxies of comparable mass have changed little since $z\approx 2.2$,
despite the observations that individual galaxies at $z\approx 2$ on
average were forming stars at $>20$ times faster rate than
low-redshift galaxies (Wuyts \etal\ 2011).  The constant spatial
profile in absorption around galaxies of disparate star formation
properties is difficult to reconcile, if these absorbers originate
primarily in starburst driven outflows.  Consequently, the origin of
halo gas revealed in absorption spectroscopy remains an open question
and to fully understand the origin and growth of gaseous halos around
galaxies requires new observations.

To go beyond the traditional one-dimensional application of QSO
probes, we have targeted two intermediate-redshift galaxies in the
field around the quadruply-lensed QSO HE\,0435$-$1223 at $z=1.689$
(Wisotzki \etal\ 2002) and searched for absorption features in the
spectra of individual lensed QSO images that are associated with the
galaxies.  The four QSO images are separated by $\approx 1.6''-2.5''$
(Figure 1) and serve as a natural integral field unit for mapping the
kinematics of halo gas around individual galaxies in the foreground.
A fundamental difference between our study and previous
absorption-line analyses toward lensed or binary QSOs (e.g.\ Rauch
\etal\ 1999, 2001a,b, 2002; Martin \etal\ 2010) is in the prior
knowledge of the locations of the associated galaxies.
Absorption-line spectroscopy of multiply-lensed QSOs near a known
foreground galaxy allows us to establish a crude three-dimensional
(3D) map of halo gas around the galaxy that records the line-of-sight
velocity field along with the two-dimensional (2D) distribution of
absorption strength.  Combining the absorption profiles revealed along
multiple sightlines with known orientation of the star-forming disk
with respect to individual sightlines allows us for the first time to
resolve the kinematics of tenuous halo gas on scales of $5-10$ kpc for
galaxies at $z>0.2$ (cf.\ Verheijen \etal\ 2007), and offers a unique
opportunity to begin to constrain models for inflows and
galactic-scale outflows.

In the field around HE\,0435$-$1223, we have identified two galaxies
in close projected distances from the lensed QSOs (at angular
separations $\theta<10''$).  As demonstrated in the following
sections, one is a typical blue star-forming galaxy at redshift
$z=0.4188$ and the other is a massive quiescent star-forming barred
spiral at $z=0.7818$.  High-quality images of the field are available
in the {\it Hubble Space Telescope} (HST) data archive, allowing us to
obtain accurate measurements of the orientation of the star-forming
disks.  Here we present spatially resolved velocity maps of halo gas
around these two intermediate-redshift galaxies based on a joint
analysis of absorption-line observations and relative alignments
between the star-forming disks and each of the four QSO sightlines.

This paper is organized as follows.  In Section 2, we describe the
observations and data reduction.  In Section 3, we present the
observed and derived photometric and spectroscopic properties of two
galaxies in the foreground of the lensed QSO.  In Section 4, we take
into account known morphologies of the galaxies and present the
velocity maps of their gaseous halos.  We compare the observations
with predictions based on different models and discuss the
implications of our results in Section 5, and summarize the results of
our study in Section 6.  We adopt a $\Lambda$CDM cosmology,
$\Omega_{\rm M}=0.3$ and $\Omega_\Lambda = 0.7$, with a Hubble
constant $H_0 = 70 \ {\rm km} \ {\rm s}^{-1}\ {\rm Mpc}^{-1}$
throughout the paper.

\begin{figure}
\begin{center}
\includegraphics[scale=0.58]{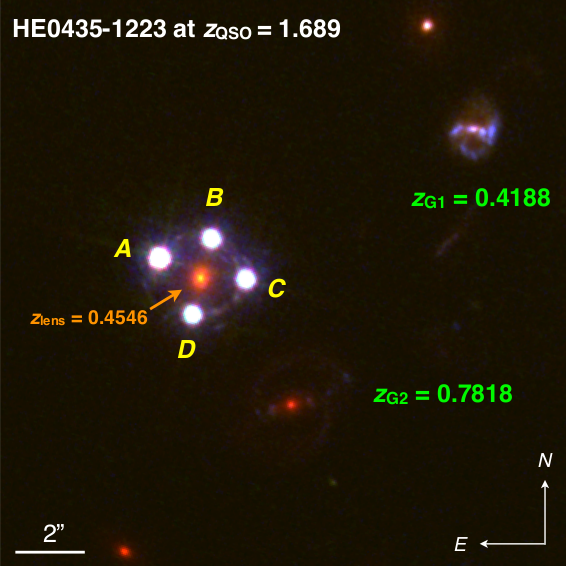}
\end{center}
\caption{Optical composite image of the field surrounding
  HE0435$-$1223 made using the HST/WFC3 UVIS channel and the F275W
  filter, and ACS WFC and the F555W and F814W filters.  The four
  lensed images ($ABCD$) of the background QSO at $z=1.689$ are well
  resolved from the lens at $z_{\rm lens}=0.4546$ with a maximum
  angular separation of $\overline{AC}=2.5''$.  Galaxies, $G1$ and
  $G2$, are spectroscopically identified at $z=0.4188$ and $z=0.7818$,
  respectively (see \S\ 3 for discussion).}
\end{figure}

\section{Observations and Data Reduction}

To spatially resolve the velocity field of halo gas around galaxies,
both high spatial resolution images of the galaxies and high spectral
resolution absorption spectra of background QSOs are needed.  The high
spatial resolution images of the galaxies allow us to determine the
orientation and inclination of the star-forming disks relative to the
lines of sight to the background QSOs.  The high spectral resolution
absorption spectra allow us to resolve individual absorbing
components, thereby characterizing gas flows in galactic halos.  In
addition, moderate-resolution spectra of the galaxies enable both
precise measurements of their systemic redshifts and accurate
measurements of the gas-phase metallicity in the star-forming ISM.
Here we describe available imaging and spectroscopic data for the
field around HE\,0435$-$1223.

\subsection{Imaging Observations}

Optical imaging observations of the field around HE\,0435$-$1223 were
performed with the HST Advanced Camera for Surveys (ACS) using the
F555W and F814W filters (PID 9744; PI: C.\ Kochanek).  Additional
imaging observations of this field were performed with the HST Wide
Field Camera 3 (WFC3) using the UVIS channel and the F275W filter (PID
11732; PI: C.\ Kochanek).  The imaging data were retrieved from the
HST data archive and processed using the standard reduction pipeline.
We registered these frames to a common origin using stars in the
field.  A color composite image of the field around the lens is
presented in Figure 1.  The mean FWHM of the point spread function for
all three filters is approximately $0.1''$.  In the ACS images, the
four lensed images ($ABCD$) of the background QSO at $z=1.689$ are
well resolved from the lens at $z_{\rm lens}=0.4546$ with a maximum
angular separation of $\overline{AC}=2.5''$ between images $A$ and
$C$.

\subsection{Galaxy Spectroscopy}

We have obtained optical spectra of two spiral galaxies, $G1$ and
$G2$, at angular separations $\theta_{\rm lens}<10''$ from the lens
(Figure 1).  The galaxies were selected based on their proximity to
the line of sight toward the quad lens.  $G1$ at $\theta_{\rm
  lens}=8.9''$ was targeted spectroscopically by Morgan \etal\ (2005),
who reported a redshift measurement of $z=0.4191\pm0.0002$.  $G2$ at
$\theta_{\rm lens}=4.4''$ was considered a possible member associated
with the gravitational lens at $z_{\rm lens}=0.45$, but detailed lens
models from previous studies by Wisotzki \etal\ (2003) and Morgan
\etal\ (2005) showed that this galaxy is likely located at a
cosmologically distinct redshift from the lens.  We targeted both
galaxies for spectroscopy.  The observations were carried out using
MagE (Marshall \etal\ 2008) on the Magellan Clay Telescope at the Las
Campanas Observatory.

MagE is a high-throughput echellette spectrograph that offers a
contiguous spectral coverage from $\lambda=3100$ \AA\ through
$1\,\mu$m.  We used a $1''$ slit and $2\times 1$ binning during
readout, which yielded a spectral resolution of ${\rm FWHM} \approx
150$ \kms.  The observations of each galaxy were carried out on the
night of 15 December 2012 in a sequence of two exposures of duration
900 s to 1800 s each.  The slit was aligned along the optimal
parallactic angle for the duration of the observations.  The mean
seeing condition over the period of integration was $0.8''$.  The
galaxy data were processed and reduced using the data reduction
software described in Chen \etal\ (2010a).  In summary, wavelengths
were calibrated using a ThAr frame obtained immediately after each
exposure and subsequently corrected to vacuum and heliocentric
wavelengths.  Cross-correlating the observed sky spectrum with the sky
emission atlas published by Hanuschik (2003) confirmed that the
wavelength calibration was accurate to $\approx 7$ \kms.  Relative
flux calibration was performed to correct for the response function of
individual echellette orders, using a sensitivity function derived
from earlier observations of the flux standard EG274.  Individual
flux-calibrated orders were coadded to form a single spectrum.

\begin{footnotesize}
\begin{table*}
\tiny
  \centering
  \begin{minipage}{160mm}
    \caption{Summary of Galaxy Properties}
    \begin{tabular}{@{}crrrcrcccccc@{}}
      \hline
      &  & \multicolumn{1}{c}{$\theta_{\rm lens}$} & \multicolumn{1}{c}{$\rho$}  &  \multicolumn{1}{c}{$i_0$}  & \multicolumn{1}{c}{P.A.$^a$} & $AB({\rm F275W}, {\rm F555W}, {\rm F814W})^b$ & $M_B$ & 12 & ${\rm EW}_{{\rm H}\alpha}$ & SFR & $M_*$  \\
      & \multicolumn{1}{c}{$z_{\rm gal}$}  & ($''$) & (kpc) & \multicolumn{1}{c}{($^{\circ}$)} & \multicolumn{1}{c}{($^{\circ}$)} & (mag) & (mag) & $+\log ({\rm O}/{\rm H})$ & (\AA) & (${\rm M}_\odot/{\rm yr}$) &  (${\rm M}_\odot$) \\
      \hline
      \hline
      $G1$  & 0.4188 & 8.9 & 49 & 40 & 173 & ($23.39\pm 0.02$, $21.16\pm 0.01$, $20.39\pm 0.01$) & $-20.5$ & $8.32\pm0.07$ & $80\pm 5$ & 4.3 & $(2-3)\times 10^{10}$\\
      $G2$  & 0.7818 & 4.4 & 33 & 25 &  13 & ($24.50\pm 0.04$, $22.35\pm 0.02$, $20.79\pm 0.01$) & $-22.0$ & ... & ... & $>0.8$ & $\approx8\times 10^{10}$ \\
      \hline
      \multicolumn{11}{l}{$^a$Position angle of the inclined disk measured north through east.}\\
      \multicolumn{11}{l}{$^b$Aperture photometry measured in a $4''$ diameter aperture centered at the galaxy.}\\
    \label{gal_table}
  \end{tabular}
\end{minipage}
\end{table*}
\end{footnotesize}

\subsection{QSO Absorption Spectroscopy}

Echellette spectroscopic observations of the lensed QSOs were obtained
using MagE on 31 August 2011.  The observations were carried out using
a $1''$ slit and $1\times 1$ binning during readout, which yielded a
spectral resolution of ${\rm FWHM} \approx 70$ \kms.  The mean seeing
condition over the period of integration was $\approx 0.7''-0.8''$.
The total exposure time accumulated for each lensed QSO ranged from
3300 seconds for images $A$ and $C$ to 3600 seconds for image $D$.
The QSO spectra were processed and reduced using the same customized
reduction pipeline described in \S\,2.2. 
Individual echellette orders were continuum normalized and coadded to
form a single spectrum that covers a spectral range from
$\lambda=3050$ \AA\ to $\lambda=1\,\mu$m.  The continuum was
determined using a low-order polynomial fit to spectral regions that
are free of strong absorption features.  The continuum-normalized
spectra have $S/N\apg 10-20$ per resolution element at $\lambda\apg
3800$ \AA.

To resolve gas kinematics in galactic halos, higher resolution
absorption spectra are necessary.  We have therefore attempted echelle
spectroscopy for the lensed QSOs, using the MIKE echelle spectrograph
(Bernstein \etal\ 2003) on the Magellan Clay telescope.  MIKE delivers
an unbinned pixel resolution of $0.12''$ along the spatial direction
and $\approx 0.02$ \AA\ along the spectral direction in the blue arm
that covers a wavelength range from $\lambda=3200$ \AA\ through
$\lambda=5000$ \AA.  The observations were carried out on the nights
of 15 and 16 of December 2012.  The mean seeing condition over this
period was $0.7''$.  Because the lensed QSOs are faint with
$g_A=19.0$, $g_B=19.5$, $g_C=19.6$, and $g_D=19.6$ mag (Wisotzki
\etal\ 2002), we experimented with two sets of heavy binning $3\times
3$ and $2\times 4$ during readout in order to achieve sufficient
signal ($S/N\approx 10$ per resolution element) in a reasonable amount
of exposure time.  While the instrument line spread function is
slightly under-sampled with the adopted binning, the observations
still allow us to resolve velocity profiles on scales of $\sim 10$
\kms.

We succeeded in observing three ($ABC$) of the four lensed QSO images
using MIKE.  QSO image $B$ was observed on the first night.  The
observations consist of $3\times 3600$ s exposures.  We used a $0.7''$
slit and $3\times 3$ binning during readout, which yielded a spectral
resolution of FWHM\,$\approx\,9$ \kms\ at $\lambda=4000$ \AA.
However, it quickly became clear that the heavy binning along the
spatial direction makes an accurate sky subtraction challenging with
image $A$ at $\approx 1.6''$ away.  We therefore observed $A$ and $C$
on the second night, using a $0.7''$ slit and $2\times 4$ (spatial
$\times$ spectral) binning during readout.  The observations include a
total exposure time of $5200$ s and 6600 s for $A$ and $C$,
respectively.

The data were processed and reduced using a customized reduction
pipeline developed by G.\ Becker and kindly offered to us by the
author.  In summary, individual spectral images were first
bias-subtracted and corrected for pixel-to-pixel variation using
twilight flats obtained through a diffuser.  Next, a 2D wavelength
map, corrected to vacuum and heliocentric wavelengths, was produced
using a ThAr frame obtained immediately after each exposure.  Object
spectra were then optimally extracted using a Gaussian profile that
matches the width of object profile along the cross-dispersion
direction in each order.  Neighboring objects (other lensed images in
the case presented here) that moved into the slit during individual
exposures were masked during spectrum extraction.  Next, flux
calibration was performed using a sensitivity function derived from
observations of the flux standard Feige\,110, and individual flux
calibrated echelle orders were coadded to form a single spectrum.
Finally, these order-combined individual exposures were continuum
normalized and stacked to form one final combined spectrum per QSO
using an optimal weighting routine.  The continuum was determined
using a low-order polynomial fit to spectral regions that are free of
strong absorption features.

\section{Analysis of Galaxy Properties}

The HST images described in \S\ 2.1 show that while the two galaxies
at $\theta_{\rm lens}<10''$ from the lens (Figure 1) can be generally
characterized as nearly face-on barred spirals, they display distinct
colors and resolved morphologies.  Specifically, $G1$ at $z=0.4188$
shows enhanced star-forming regions with luminous UV radiation across
the central bar and along the spiral arms, and $G2$ at $z=0.7818$
shows a well-developed central bar and two spiral arms that are
dominated by older stellar population with some trace of on-going star
formation at the ends of the bar and the spiral arms (as revealed by
faint UV emission; see also Figure 7).

We measure the inclination angle ($i_0$) of each galaxy based on the
ratio of the observed spatial extent along the major and minor axes,
and find that $i_0= 40^\circ$ for $G1$ and $i_0= 25^\circ$ for $G2$.
Uncertainties in the inclination angle are roughly $\pm 3^\circ$.  We
also measure the position angle of the major axis on the sky and find
${\rm P.A.}= 173^\circ$ measured north through east for $G1$ and ${\rm
  P.A.}= 13^\circ$ for $G2$.  Uncertainties in the position angle are
roughly $\pm 5^\circ$.  Here we summarize photometric and
spectroscopic properties of these two galaxies.

\begin{figure}
\begin{center}
\includegraphics[scale=0.375]{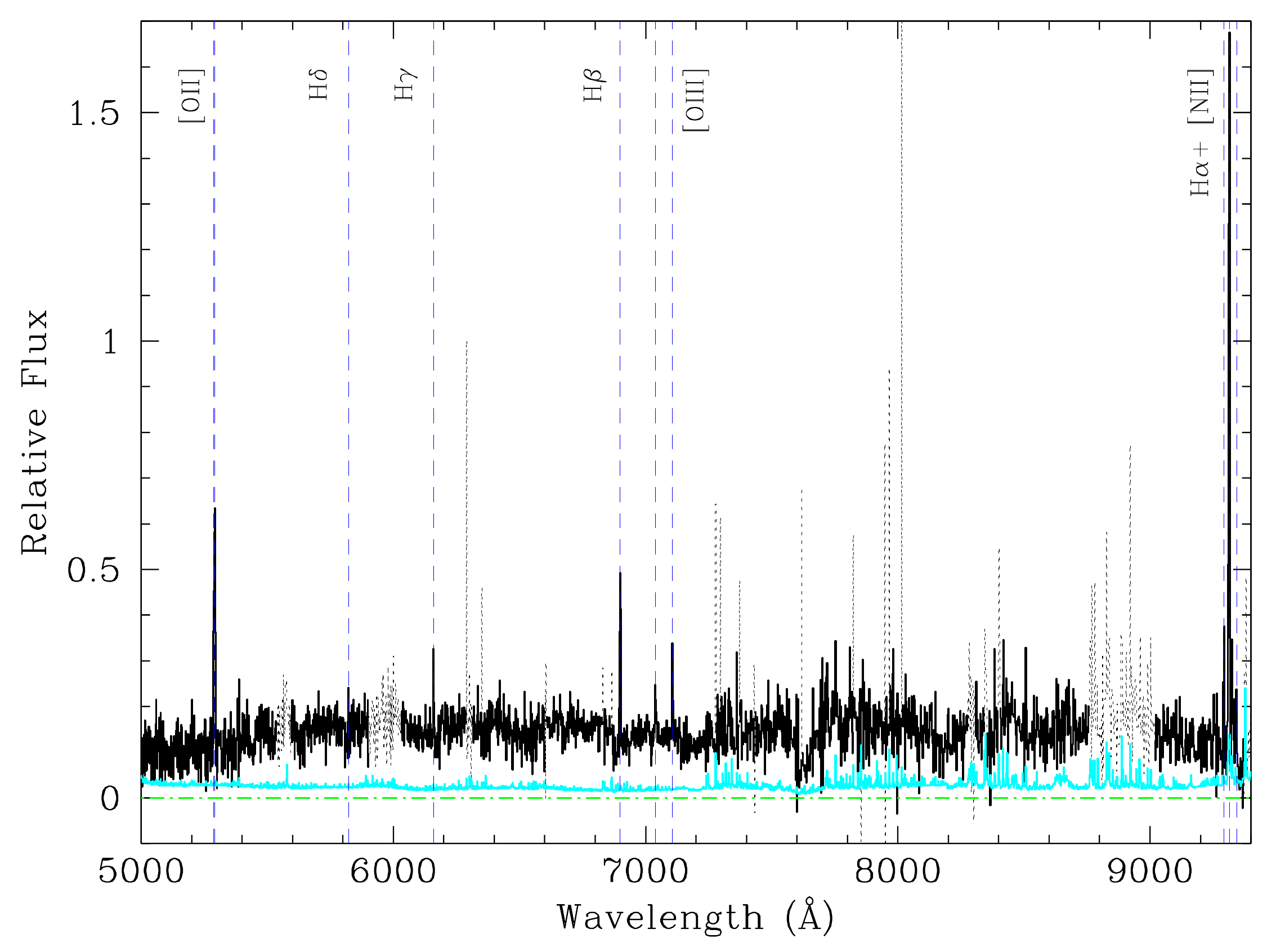}
\end{center}
\caption{Optical spectrum of $G1$ obtained using MagE on the Magellan
  Clay Telescope.  The data reveal numerous emission features (marked
  by long dashed lines) with a mean redshift at $z=0.4188\pm 0.0001$.
  The corresponding 1-$\sigma$ error spectrum is shown in cyan at the
  bottom above the zero flux level (dash-dotted line).  Contaminating
  sky residuals are dotted out for clarification.}
\end{figure}

\subsection{$G1$ at $z=0.4188$}

We first note that Morgan \etal\ (2005) analyzed the ACS F555W and
F814W images and measured $4''$-diameter aperture photometry for this
galaxy.  These authors reported $m_{\rm F555W}=21.18\pm 0.02$ and
$m_{\rm F814W}=21.16\pm 0.01$ in the STmag system (e.g.\ Koorneef
\etal\ 1986).  In addition, they obtained an optical spectrum of the
galaxy and measured a mean redshift $z=0.4191\pm 0.0002$ based on
observations of [O\,II], H$\beta$, and [O\,III].  Furthermore, these
authors also noted the presence of a second galaxy at $z=0.4189$ at
3.5 times angular distance away, $\theta_{\rm lens}=31.4''$ (G22 in
their paper), suggesting that $G1$ may be associated with a galaxy
group.

Our MagE observations of $G1$ confirm that the galaxy spectrum is
dominated by a blue continuum and strong emission line features
(Figure 2).  In addition to [O\,II], H$\beta$ (which occurs in the
terrestrial atmospheric B band absorption), and [O\,III], we also
detect [N\,II], H$\alpha$, and higher-order Balmer emission lines.  A
cross-correlation analysis with a linear combination of SDSS eigen
spectra of galaxies returns a best-fit redshift and redshift
uncertainty of $z=0.4188\pm 0.0001$.  At this redshift, the projected
distance between $G1$ and the lens is $\rho=49$ kpc.

We determine the interstellar oxygen abundance of $G1$ using the
semi-empirical correlation between $12 + \log({\rm O}/{\rm H})$ and
$N2\equiv\log\,[{\rm N\,II}]/{\rm H}\alpha$ from Pettini \& Pagel
(2004).  We adopt the $N2$ index for the interstellar oxygen abundance
measurement of $G1$, because it is based on two closely located lines
([N\,II]\,$\lambda\,6853$ and H$\alpha$) and not sensitive to
uncertainties in flux calibration and/or differential dust extinction.
Based on the observed [N\,II] and H$\alpha$ line ratio, we derive $12
+ \log({\rm O}/{\rm H}) = 8.32 \pm 0.07$ for the gas phase abundance
in the ISM of $G1$.

To determine the intrinsic luminosity, color, stellar mass, and star
formation rate (SFR) of the galaxy, we also measure aperture
photometry of galaxy $G1$ in the archival optical and UV images
obtained using HST ACS and WFC3.  Within a $4''$ diameter aperture, we
measure $AB({\rm F275W})=23.39\pm 0.02$, $AB({\rm F555W})=21.16\pm
0.01$, $AB({\rm F814W})=20.39\pm 0.01$.  At $z=0.4188$, the observed
apparent magnitudes lead to rest-frame absolute magnitudes of
$M_B=-20.5$ in the $B$ band and $M_R=-21.3$ in the $R$ band, and $g-r$
color of $g-r=0.4$ mag for the galaxy.  

Adopting the color-based stellar mass-to-light ratio $M_*/L$ from Bell
\etal\ (2003) for different stellar initial mass functions, we
estimate a total stellar mass of $M_*\approx (2-3)\times 10^{10}\,{\rm
  M}_\odot$ for $G1$.  Adopting the stellar mass to halo mass relation
of Behroozi \etal\ (2010), we further estimate the dark matter halo
mass of $G1$ to be $M_h\sim 10^{12}\,{\rm M}_\odot$.  Finally, we
estimate the on-going SFR based on the observed H$\alpha$ equilvalent
width (${\rm EW}_{{\rm H}\alpha}$) and rest-frame $R$-band magnitdue.
We measure ${\rm EW}_{{\rm H}\alpha}=-80\pm 5$ \AA\ and infer a total
H$\alpha$ flux of $f_{{\rm H}\alpha}=8\times 10^{41}\ {\rm erg}\,{\rm
  s}^{-1}$ under the assumption that ${\rm EW}_{{\rm H}\alpha}$ is
roughly constant across the disk.  Adopting the star formation rate
calibration of Kennicutt \& Evans (2012), we estimate an unobscured
SFR of $\approx 4.3\ {\rm M}_\odot\,{\rm yr}^{-1}$ for $G1$ which is
roughly six times higher than the SFR inferred from the observed UV
flux in the F275W band.  The difference between H$\alpha$ and UV
inferred SFR can be attributed to either dust extinction or spatial
inhomogeneities in ${\rm EW}_{{\rm H}\alpha}$.  We cannot constrain
the dust content based on the observed flux ration between H$\alpha$
and H$\beta$ lines, because H$\beta$ falls in the atmosphere $B$-band
absorption.

A summary of the optical properties of $G1$ is presented in the first
row of Table 1.  We conclude that $G1$ at $z=0.4188$ is an $L_*$
galaxy with photometric properties consistent with typical blue
star-forming galaxies at $z\sim 0.4$ (e.g.\ Noeske \etal\ 2007; Zhu
\etal\ 2011).

\begin{figure}
\begin{center}
\includegraphics[scale=0.275]{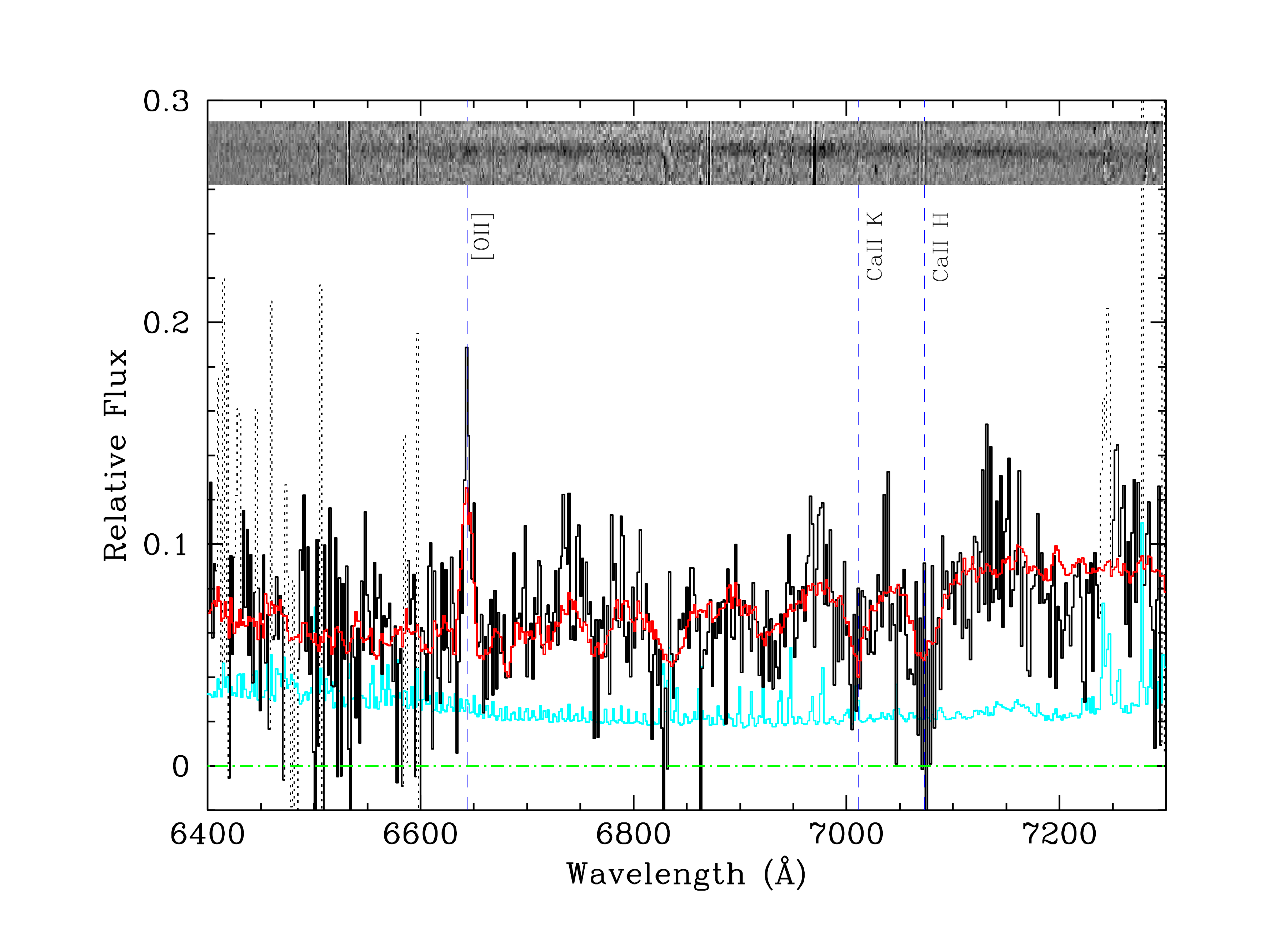}
\end{center}
\caption{Optical spectrum of $G2$ obtained using MagE on the Magellan
  Clay Telescope.  The corresponding 1-$\sigma$ error spectrum is
  shown in cyan at the bottom above the zero flux level (dash-dotted
  line), while a rectified two-dimensional spectral image is presented
  at the top.  Contaminating sky residuals are dotted out for
  clarification.  We identify a faint emission line at 6644 \AA, along
  with absorption features that are consistent with [O\,II] and Ca\,II
  H\&K absorption doublet at $z=0.7818$.  We measure the redshift
  using a cross-correlation analysis with a linear combination of SDSS
  eigen spectra of galaxies.  The best-fit model template is shown as
  the red spectrum, which also displays matched absorption features
  due to the Balmer series.}
\end{figure}

\subsection{$G2$ at $z=0.7818$}

Optical photometry of this galaxy has also been reported by Morgan
\etal\ (2005), who measured $m_{\rm F555W}=22.25\pm 0.04$ and $m_{\rm
  F814W}=21.26\pm 0.01$ in the STmag system over a $4''$-diameter
aperture.  However, no spectroscopic observations have been made by
this team.  It is possible that $G2$ and the lens are correlated,
because of a small angular distance ($\theta_{\rm lens}=4.4''$,
corresponding to $\rho=25$ kpc at $z=0.45$) between the two galaxies.
But as noted earlier by Wisotzki \etal\ (2003) and Morgan \etal\
(2005), detailed lens models have shown that $G2$ is more likely
located at a cosmologically distinct redshift from the lens.

Our MagE observations of $G2$ revealed a faint emission line at 6644
\AA, along with absorption features that are consistent with [O\,II]
and Ca\,II H\&K absorption doublet at $z=0.7818$.  Figure 3 shows both
the extracted one-dimensional spectrum and the corresponding
two-dimensional spectral image.  At this redshift, H$\beta$ and
[O\,III] are covered by the MagE data but occur in the forest of OH
sky lines.  We therefore cannot robustly determine the
presence/absence of H$\beta$ and [O\,III] lines.  A cross-correlation
analysis with a linear combination of SDSS eigen spectra of galaxies
returns a best-fit redshift and redshift uncertainty of $z=0.7818\pm
0.0004$.  We present the best-fit model spectrum in red in Figure 3.
Comparing the observed spectrum and best-fit model also reveals the
presence of Balmer absorption series, suggesting a post-starburst
nature of the galaxy.

We have also measured aperture photometry of galaxy $G2$ in the
archival optical and UV images obtained using HST ACS and WFC3.
Within a $4''$ diameter aperture, we measure $AB({\rm F275W})=24.50\pm
0.04$, $AB({\rm F555W})=22.35\pm 0.02$, $AB({\rm F814W})=20.79\pm
0.01$.  At $z=0.7818$, the observed F814W band corresponds well with
rest-frame $B$ band.  The observed F814W magnitude therefore
translates to a rest-frame $B$-band absolute magnitude of $M_B=-22.0$
for $G2$, which is roughly $1.7\,L_*$ (Faber \etal\ 2007).  

At $z=0.78$, the rest-frame $R$ band is redshifted into the observed
$J$ band.  Without near-infrared photometry, we cannot accurately
determine the total stellar mass of $G2$.  While the apparent red
color of $G2$ in Figure 1 indicate little on-going star formation, an
accurate estimate of the total stellar would require rest-frame
optical and near-infrared colors.  Adopting the same
mass-to-light ratio from $G1$ leads to $M_*\approx 8\times
10^{10}\,{\rm M}_\odot$ and $M_h\sim 5\times 10^{12}\,{\rm M}_\odot$
for $G2$.  Finally, we constrain the on-going SFR based on the
observed UV flux in the F275W band and derive an unobscured SFR of $>
0.8\ {\rm M}_\odot\,{\rm yr}^{-1}$ for $G2$.

A summary of the optical properties of $G2$ is presented in the second
row of Table 1.  We conclude that $G2$ at $z=0.7818$ is a massive,
super-$L_*$ galaxy with spectroscopic properties similiar to quiescent
star-forming galaxies at $z=0.5-1$ (e.g.\ Poggianti \etal\ 2009).

\section{Velocity Maps of Gaseous Halos}

The available HST images have yielded important constraints for both
the photometric and morphological properties of galaxies $G1$ and
$G2$.  The next step is to analyze the absorption profiles revealed
along multiple sightlines, in order to establish a spatially resolved
velocity map of absorbing clouds in the galactic halo.

The absorption-line analysis is, however, complicated due to a mixed
spectral quality in the available data.  As described in \S\ 2.3, MIKE
echelle spectroscopy was carried out for QSO sightlines $A$, $B$, and
$C$, while MagE echellette observations were carried out for QSO
sightlines $A$, $C$, and $D$ during two different observing periods.
The native spectral resolutions offered by MIKE and MagE are $\approx
10$ and $\approx 70$ \kms, respectively.  In addition, because the QSO
images are fainter than the nominal source brightness for attempting
MIKE spectroscopy, we adopted heavy binning during readout to increase
the observation efficiency.  As a result, we are able to separate
individual absorbing components separated by $\apg 10$ \kms\ in the
MIKE spectra, but we are unable to resolve individual resolution
elements.  For absorption spectra obtained with MagE, we cannot
resolve features on scales $<70$ \kms.  The expectation is that while
fluctuations on small velocity scales along a single line of sight may
be lost due to heavy binning or low spectral resolution, large-scale
velocity gradient across different lines of sight can still be
resolved.

In this section, we first describe the procedures developed for
analyzing the under-sampled absorption spectra.  The results from
individual sightlines are then combined to establish a crude 3D
velocity map for the gaseous halos around $G1$ and $G2$, which are
described in the subsequent sections.

\subsection{Analysis of Absorption Profiles}

A Mg\,II absorber is detected in every sightline through the halo of
each of the two galaxies in our study.  The absorber strength observed
in these galaxies is typical of what is seen in previous close
QSO--galaxy pair studies (e.g.\ Chen \etal\ 2010a,b).  We focus our
analysis on the observed Mg\,II absorption doublet, because these are
the strongest transitions that can be observed in the optical echelle
data and because in nearly all four sightlines these are the only
transitions that are detected at $>10\,\sigma$ level of significance.
Given that only Mg\,II absorption is observed, we cannot obtain direct
constraints for the physical properties of the gas.  In the following
analysis, we infer a mean gas metallicity based on the empirical
metallicity--$W_r(2796)$ relation of Murphy \etal\ (2007; see also
Ledoux \etal\ 2006), and assume an ionization fraction of $f_{\rm
  Mg^+}=0.1$ based on a simple ionization model presented in Chen \&
Tinker (2008).

For $G1$, we have MIKE spectra available for sightlines $A$, $B$, and
$C$ and MagE spectra available for $D$ for a complete four-point
mapping of its gaseous halo as offered by the quad-lens system.  For
$G2$, however, the associated Mg\,II absorption doublet occur near the
cross-over wavelength of the dichroic in MIKE, where the
signal-to-noise ($S/N$) of the data is particularly poor.
Consequently, only MagE spectra of $A$, $C$, and $D$ are available for
mapping the gaseous halo of $G2$.

To account for the mixed spectral quality and derive a robust velocity
map, we develop custom computer programs for analyzing the observed
absorption profiles.  The primary goal of our analysis is to obtain an
accurate map of the velocity field across different sightlines.

We first characterize the observed absorption along each sightline as
due to discrete clouds.  Under this scenario, we generate a model
absorption profile based on the sum of a minimum number, $n_c$, of
Voigt profiles necessary to explain the observed kinematic signatures.
The model depends on a set of free parameters, including for each
component the velocity offset with respect to the systemic redshift of
the galaxy ($\Delta\,v_c$), the Mg\,II absorption column density,
$\log\,N_c({\rm Mg\,II})$, and Doppler parameter, $b_c$, necessary to
define the Voigt profile.  We then convolve the model profile with a
Gaussian function that is appropriate for simulating the instrument
resolution offered by MIKE/MagE.  We bin the convolved model spectrum
to match the size of the spectral pixel in the data.  We then compare
the binned model spectrum with the observed absorption spectrum and
its corresponding error spectrum, and calculate the $\chi^2$.
Finally, we constrain the model parameters by minimizing the $\chi^2$.

We note that because the absorption spectra are under-sampled and many
components are saturated (see Figure 4 for examples), $N_c({\rm
  Mg\,II})$ and $b_c$ are not well constrained and serve only as a
guide.  On the other hand, the velocity centroid of each identified
component is well determined to be better than a small fraction of the
resolution element ($\sim 2$ \kms), which is the key measurement
necessary for constructing an accurate 3D velocity map.

Next, we characterize the observed absorption as due to density
fluctuations in the Mg$^+$ ions and measure the effective optical
depth $\tau_{\rm eff}$ as a function of line-of-sight velocity offset
$\Delta\,v$.  The effective optical depth is defined as
\begin{equation}
\tau_{\rm eff}=-\ln\,\langle\,F\,\rangle\equiv -\ln\,\langle\,e^{-\tau}\,\rangle,
\end{equation}
where $\tau$ is the underlying Mg\,II optical depth and
$\langle\,F\,\rangle$ is the mean continuum-normalized flux per pixel.
We compute $\tau_{\rm eff}$ using the best-fit model from the Voigt
profile analysis described above to avoid saturated pixels or pixels
with negative fluxes.  The observed $\tau_{\rm eff}$ provides a simple
and non-parametric characterization of how the mean Mg\,II absorption
varies with $\Delta\,v$ along individual sightlines.  It does not
depend on detailed assumptions of the number of absorption clumps that
may be blended in each observed absorption component.  We also compute
the optical depth-weighted mean velocity offset, $\Delta\,v_\tau$, and
the velocity width that encloses 90\% of the total line-of-sight
effective optical depth, $\delta\,v_{90}$, for quantifying the
line-of-sight gas kinematics.  Because $\Delta\,v_\tau$ is calculated
based on the effective optical depth per pixel, the uncertainties in
$\Delta\,v_\tau$ is approximately the pixel resolution of the data,
which is 10 \kms\ for MIKE and 40 \kms\ for MagE spectra.

\subsection{Spatially Resolved Halo Gas Kinematics around the Blue
  Star-forming Galaxy $G1$ at $z=0.4188$}

Absorption kinematics of halo gas around the $L_*$ galaxy $G1$ is
displayed in Figure 4, which shows the Mg\,II $\lambda\lambda$
2796,2803 absorption profiles observed along the four lensed QSO
sightlines.  The zero relative velocity in each panel corresponds to
the systemic redshift of the galaxy at $z=0.4188$.  The best-fit model
spectrum of each sightline (obtained following the procedures
described in \S\ 4.1) is also presented in red for comparison.  The
observations have uncovered a relatively uniform coverage of Mg\,II
absorbing gas at $\sim 50$ projected kpc from the star-forming disk,
with all four sightlines displaying a strong Mg\,II absorber of
$W_r(2796)\apg 1$ \AA.  In addition, while individual components along
individual sightlines are observed to spread over a large velocity
range, $\sim 160$ \kms\ along the sightline toward the $B$ image and
$\sim 140$ \kms\ toward $C$, the absorption morphology appears to be
remarkably similar across these sightlines with a dominant absorption
pair occuring near the systemic redshift of $G1$ and trailed by a
second pair of absorption features at $\approx +110$ \kms.  Only a
relatively small velocity shear ($\Delta\,v\approx 20$ \kms) is seen
between different sightlines that are separated by $8-10$ kpc in
projected distances.

\begin{center}
\begin{table*}
    \caption{Spatial Variation of Mg\,II Absorption Properties$^a$ around $G1$ at $z=0.4188$}
    \begin{tabular}{cccccccccrcc}
      \hline
      &  \multicolumn{1}{c}{$\theta_{\rm qim}$} & \multicolumn{1}{c}{$\rho$}  &  \multicolumn{1}{c}{$\alpha^c$} & \multicolumn{1}{c}{$W_r(2796)$}  & \multicolumn{1}{c}{$\Delta\,v_\tau$} & \multicolumn{1}{c}{$\delta\,v_{90}$} & & & \multicolumn{1}{c}{$\Delta\,v_c$} &  & \multicolumn{1}{c}{$b_c$} \\
      Sightline$^b$  & ($''$)  & (kpc) &  ($^\circ$) & (\AA) & \multicolumn{1}{c}{(\kms)} & \multicolumn{1}{c}{(\kms)} & \multicolumn{1}{c}{$n_c$} & component & \multicolumn{1}{c}{(\kms)} &  $\log\,N_c({\rm Mg\,II})$ & \multicolumn{1}{c}{(\kms)} \\
      \hline
      \hline
      $A$  & 9.7 & 53.6 & 119 & $1.1\pm 0.1$ & $+76$ & 169 & 3 & 1 &  $+26$ & 13.1 & 19 \\
      &     &      &     &              &       &     &   & 2 &  $+64$ & 13.4 & 19 \\
      &     &      &     &              &       &     &   & 3 & $+153$ & 13.1 & 13 \\
      \hline
      $B$  & 8.1 & 44.7 & 120 & $1.9\pm 0.1$ & $+32$ & 185 & 4 & 1 &  $-23$ & 13.7 & 27 \\
      &     &      &     &              &       &     &   & 2 &  $+19$ & 14.0 & 27 \\
      &     &      &     &              &       &     &   & 3 &  $+96$ & 13.2 & 27 \\
      &     &      &     &              &       &     &   & 4 & $+135$ & 13.5 & 27 \\
      \hline
      $C$  & 7.8 & 43.1 & 130 & $1.5\pm 0.1$ & $+18$ & 163 & 4 & 1 &  $-11$ & 14.2 & 15 \\
      &     &      &     &              &       &     &   & 2 &  $+41$ & 13.4 & 15 \\
      &     &      &     &              &       &     &   & 3 &  $+95$ & 12.8 & 20 \\
      &     &      &     &              &       &     &   & 4 & $+125$ & 12.5 & 20 \\
      \hline
      $D$  & 9.6 & 53.0 & 130 & $1.0\pm 0.1$ & $+54$ & 161 & 2 & 1 &  $+22$ & 13.6 & 22 \\ 
      &     &      &     &              &       &     &   & 2 & $+112$ & 13.1 & 22 \\
      \hline
      \multicolumn{12}{l}{$^a$Due to a heavy binning in the data, $N_c({\rm Mg\,II})$ and $b_c$ are not well constrained in the Voigt profile analysis and serve only as a guide.} \\
      \multicolumn{12}{l}{$^b$The projected distances between different sightlines are $(\overline{\rm AB}, \overline{\rm AC}, \overline{\rm AD}, \overline{\rm BC}, \overline{\rm BD}, \overline{\rm CD})=(8.8, 13.8, 10.3, 8.2, 12.2, 10.0)$ kpc.} \\
      \multicolumn{12}{l}{$^c$Azimuthal angle of the QSO sightline from the major axis of the star-forming disk (measured counterclockwise).}\\
    \label{g1_table}
  \end{tabular}
\end{table*}
\end{center}

As described in \S\ 3.1, $G1$ is a typical blue star-forming galaxy at
$z=0.4188$ and $\sim 50$ kpc in projected distance from the quad-lens
system.  The galaxy has a fainter companion at $\approx 120$ kpc in
projected distance {\it farther} away from the QSO sightlines.  The
star-forming disk of $G1$ is characterized by an inclination angle of
$i_0=40^\circ$ and position angle of ${\rm P.A.}=173^\circ$ measured
north through east.  In Table 2, we list for each sightline the
angular separation between $G1$ and each QSO image $\theta_{\rm qim}$,
the corresponding projected distance $\rho$, the azimuthal angle
$\alpha$ of the QSO sightline with respect to the major axis of the
star-forming disk, the rest-frame absorption equivalent width
intergrated over all components $W_r(2796)$, the optical
depth-weighted mean velocity offset $\Delta\,v_\tau$, the velocity
width that encloses 90\% of the total line-of-sight effective optical
depth, $\delta\,v_{90}$, the number of components $n_c$ necessary to
reproduce the observed absorption profile along each sightline, and
the respective best-fit Voigt profile parameters of each Mg\,II
absorbing component [$\Delta\,v_c$, $\log\,N({\rm Mg\,II})$, $b_c$].
Given the observed $W_r(2796)$, we further estimate a chemical
enrichment level of $f_{Z}=0.1-0.4$ solar metallicity for the Mg\,II
absorbers around $G1$ based on the redshift-corrected
metallicity--$W_r(2796)$ relation of Murphy \etal\ (2007).  Including
the scatter observed by Murphy et al.\ (2007), the gas metallicity can
be as high as 0.9 solar and as low as 0.05 solar.

\begin{figure}
\begin{center}
\includegraphics[scale=0.45]{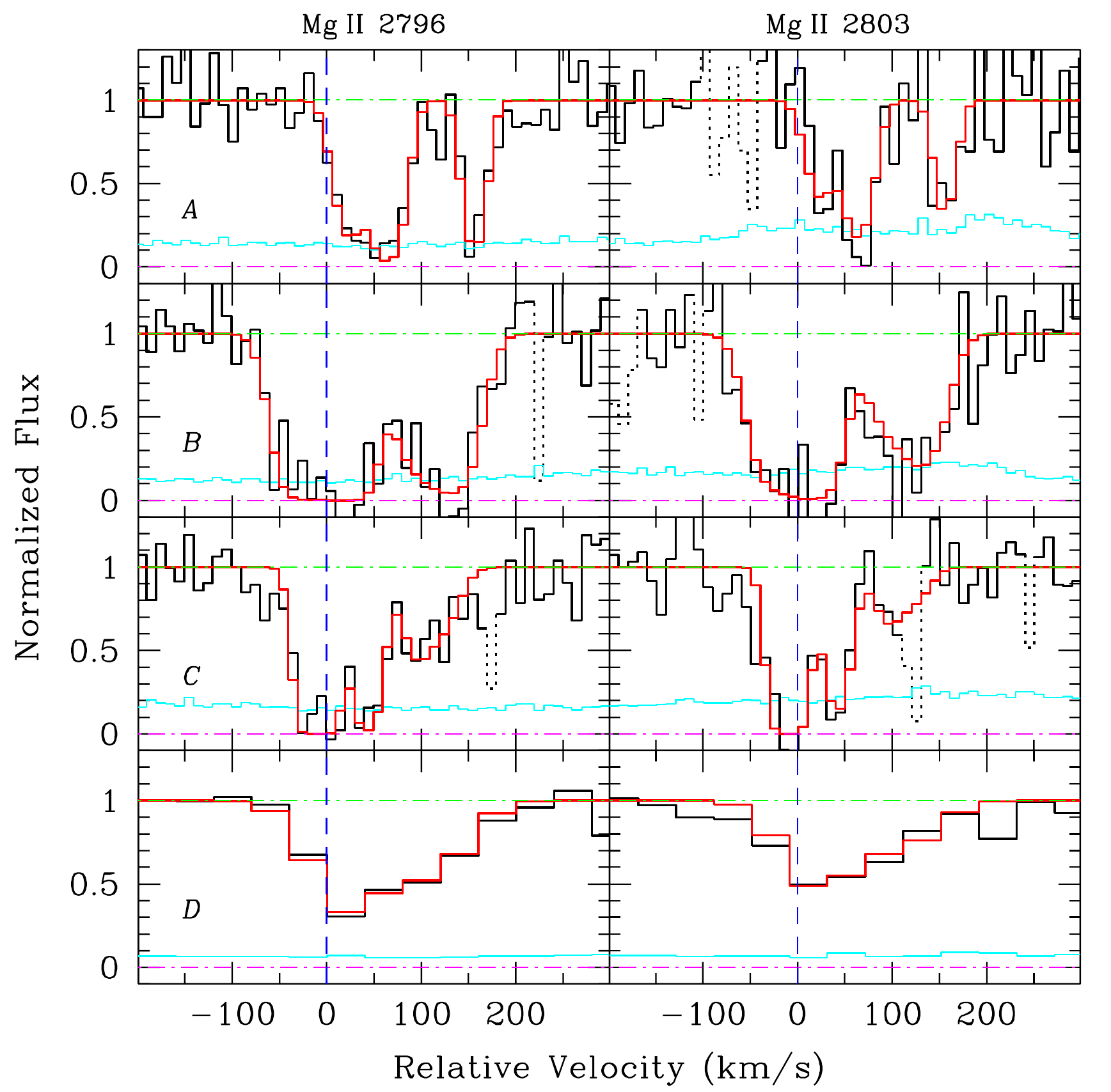}
\end{center}
\caption{Mg\,II $\lambda\lambda$ 2796,2803 absorption profiles along
  four different sightlines through the halo around the $L_*$ galaxy
  $G1$ at $z=0.4188$.  Zero relative velocity corresponds to the
  systemic redshift of the galaxy.  In every panel, the continuum
  normalized absorption spectrum is shown in black solid histograms
  with contaminating features dotted out for clarification, and the
  corresponding 1-$\sigma$ error spectrum is shown in thin cyan.
  The absorption spectra of $A$, $B$, and $C$ are from the
  high-resolution MIKE observations, while the spectrum of $D$ is from
  MagE (\S\,2.3).  The red spectrum shows the best-fit model (with a
  reduced $\chi^2\approx 2$), which takes into account the instrument
  resolution and pixel binning during readout (\S\,4.1).  The Mg\,II
  absorption doublet is generally characterized by a dominant
  component near the systemic velocity, which is followed by secondary
  absorbing components at $\sim 100$ \kms\ in the red.  Such kinematic
  signatures apply to all four sightlines separated by $8-10$ kpc in
  projected distances, and only a relatively small velocity shear
  (between $\Delta\,v\approx 20$ \kms) is seen across these different
  sightlines.}
\end{figure}

\begin{figure}
\begin{center}
\includegraphics[scale=0.45]{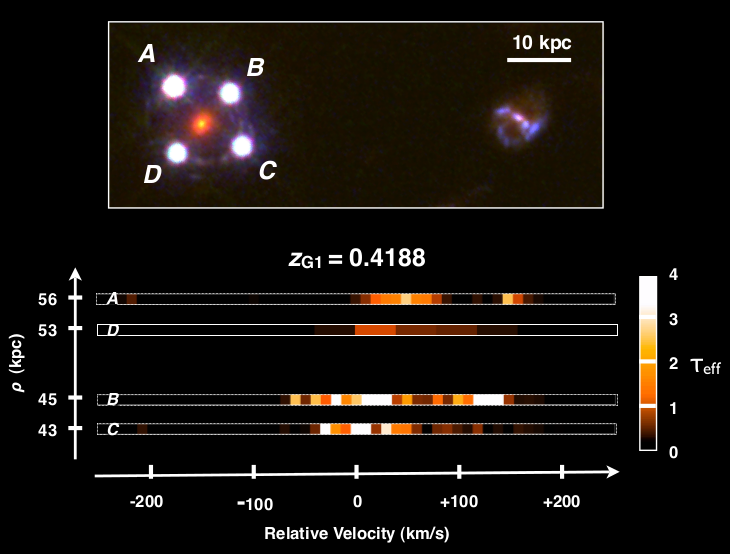}
\end{center}
\caption{Spatially resolved halo gas kinematics around the $L_*$
  galaxy $G1$ at $z=0.4188$.  The top panel shows the relative
  orientation of the quadruply-lensed QSO to $G1$ at $\rho \sim 50$
  kpc.  The spectral images at the bottom display the effective
  optical depth $\tau_{\rm eff}$ of Mg\,II versus line-of-sight
  velocity offset from $z=0.4188$ (the systemic redshift of $G1$) for
  individual sightlines with increasing projected distance $\rho$ from
  bottom to top.  The values of $\tau_{\rm eff}$ are indicated by the
  color bar in the lower-left corner.  Note that the projected
  distances between any two sightlines are very similar, $\sim 10$
  kpc.  The apparent velocity shear is found to be $\sim 40$ \kms\
  between $A$ and $B$ or between $C$ and $D$ with increasing $\rho$,
  while a velocity difference of $\apl 20$ \kms\ is seen between $A$
  and $D$ or between $B$ and $C$ that are at similar projected
  distances.}
\end{figure}

Combining known orientation of the quadruply-lensed QSOs to $G1$ from
available HST images with the observed Mg\,II $\lambda\lambda$
2796,2803 absorption profiles along individual QSO sightlines leads a
3D velocity map of halo gas around the galaxy.  The top panel of
Figure 5 shows the relative orientation of the quadruply-lensed QSO to
$G1$ at $\rho \sim 50$ kpc.  The spectral images in the bottom of
Figure 5 display the effective optical depth $\tau_{\rm eff}$ of
Mg\,II versus line-of-sight velocity offset from $z=0.4188$ (the
systemic redshift of $G1$).  The values of $\tau_{\rm eff}$ are
indicated by the color bar in the lower-left corner.  We note that the
projected distances between different sightlines are $(\overline{\rm
  AB}, \overline{\rm AC}, \overline{\rm AD}, \overline{\rm BC},
\overline{\rm BD}, \overline{\rm CD})=(8.8, 13.8, 10.3, 8.2, 12.2,
10.0)$ kpc.  From the inner sightlines $B$ and $C$ at $\rho \approx
45$ kpc (bottom two spectral images) to the outer ones $A$ and $D$ at
$\rho\approx 55$ kpc (top two spectral images), we observe a
consistent increase in the relative velocity offsets of regions where
dominant absorption occurs.  The spatial coherence between four
different sightlines, as evident from both the absorption signatures
presented in Figure 4 and the velocity shears displayed in Figure 5,
strongly suggests that the absorbing gas follows an organized motion.

\subsection{Spatially Resolved Halo Gas Kinematics around the
  Quiescent Star-forming Galaxy $G2$ at $z=0.7818$}

Absorption kinematics of halo gas around the super-$L_*$ galaxy $G2$
is displayed in Figure 6, which shows the Mg\,II $\lambda\lambda$
2796,2803 absorption profiles observed along three 
of the four QSO sightlines.  The zero relative velocity in each panel
corresponds to the systemic redshift of the galaxy.  Recall that the
Mg\,II absorption doublet associated with $G2$ occur near the
cross-over wavelength of the dichroic in MIKE where the $S/N$ of the
spectra is particularly poor.  Only moderate-resolution MagE spectra
are available for detecting the Mg\,II absorption feature at this
redshift.  In every panel, the continuum normalized absorption
spectrum is shown in black solid histograms with contaminating
features dotted out and the corresponding 1-$\sigma$ error spectrum
shown in thin cyan.  The best-fit model spectrum of each sightline
that takes into account the spectral resolution of the instrument as
described in \S\ 4.1 is also presented in red for comparison.

Similar to $G1$, the observations have uncovered a relatively uniform
coverage of Mg\,II absorbing gas at $\sim 30$ projected kpc from the
barred spiral, with all three sightlines showing the presence of
Mg\,II absorption.  In contrast to $G1$, however, the gaseous halo
around $G2$ displays only a moderately strong Mg\,II absorber of
$W_r(2796)= 0.5-0.7$ \AA\ along these sightlines at smaller projected
distances from the galaxy.  The observed $W_r(2796)$ implies a low
chemical enrichment level of $f_{Z}=0.06$ solar metallicity for the
Mg\,II absorbers around $G2$ based on the redshift-corrected
metallicity--$W_r(2796)$ relation of Murphy \etal\ (2007).  Including
the scatter observed by Murphy et al.\ (2007), the gas metallicity can
be as high as 0.15 solar and as low as 0.01 solar.  In addition,
despite a lower spectral resolution, the MagE data show remarkably
distinct absorption morphology across different sightlines.  In
particular for $C'$ and $D'$ (after lensing correction at $z=0.7818$;
see the image in Figure 7) at a similar projected distance from $G2$
and roughly 6 kpc apart, the observed Mg\,II absorption is spread over
130 \kms\ along the sightline toward $C'$, whereas more concentrated
absorption is found within a narrow ${\rm FWHM}\approx 42$ \kms\ along
the sightline toward $D'$.

As described in \S\ 3.2, $G2$ is a massive, quiescent star-forming
galaxy at $z=0.7818$ and $\sim 30$ kpc in projected distance from the
quad-lens system.  No companion is known for this galaxy.  The
star-forming disk of $G2$ is nearly face-on with an inclination angle
of $i_0=25^\circ$ and position angle of ${\rm P.A.}=13^\circ$ measured
north through east.  In addition, $G2$ occurs behind the lensing
galaxy at $z_{\rm lens}=0.4546$ and the angular separations observed
between lensed QSO images only apply to objects at $z\apl z_{\rm
  lens}$.  Assuming that the QSO is located directly behind the
lens\footnote{We note that Morgan et al.\ (2005) find that a lens
  model should include shear from neighboring galaxies in order to
  simultaneously reproduce the observed locations {\it and} flux
  ratios of the lensed quasar.  With shear included, the unknown
  source position is found by the best-fit model to be $0.14''$W and
  $0.13''$S from the center of the lens (The best-fit parameters were
  generously provided to us by C.\ Kochanek, private communication).
  Assuming this source position, the impact parameters we report in
  Table 3 and \S\,4.3 change by less than 1 kpc, which does not affect
  the results reported in this paper.}, we compute the lensed image
positions at the $z=0.7818$ plane through ray tracing.  The results
are shown as green stars in the top panel of Figure 7.

\begin{center}
\begin{table*}
    \caption{Spatial Variation of Mg\,II Absorption Properties$^a$ around $G2$ at $z=0.7818$}
    \begin{tabular}{@{}cccrcrrccrcc@{}}
      \hline
      &  \multicolumn{1}{c}{$\theta_{\rm qim}$} & \multicolumn{1}{c}{$\rho$}  &  \multicolumn{1}{c}{$\alpha^c$} & \multicolumn{1}{c}{$W_r(2796)$}  & \multicolumn{1}{c}{$\Delta\,v_\tau$} & \multicolumn{1}{c}{$\delta\,v_{90}$} & & & \multicolumn{1}{c}{$\Delta\,v_c$} & & \multicolumn{1}{c}{$b_c$} \\
      Sightline$^b$  & ($''$)  & (kpc) &  \multicolumn{1}{c}{($^\circ$)} & (\AA) & \multicolumn{1}{c}{(\kms)} & \multicolumn{1}{c}{(\kms)} & \multicolumn{1}{c}{$n_c$} & component & \multicolumn{1}{c}{(\kms)} & $\log\,N_c({\rm Mg\,II})$ & \multicolumn{1}{c}{(\kms)} \\
      \hline
      \hline
     $A'$  & 4.9 & 36.8 & 26 & $0.51\pm 0.03$ & $+187$ & 280 & 2 & 1 & $+123$ & 12.8 & 91  \\
           &     &      &    &                &        &     &   & 2 & $+258$ & 12.8 & 23 \\
     $C'$  & 4.1 & 31.0 & 16 & $0.72\pm 0.03$ &  $+97$ & 280 & 2 & 1 &  $+57$ & 13.2 & 61 \\
           &     &      &    &                &        &     &   & 2 & $+185$ & 12.9 & 20 \\
     $D'$  & 4.2 & 31.1 & 27 & $0.56\pm 0.04$ &  $+36$ & 102 & 1 & 1 &  $+35$ & 13.3 & 25  \\ 
     \hline
      \multicolumn{12}{l}{$^a$With a moderate spectral resolution, $N_c({\rm Mg\,II})$ and $b_c$ are not well constrained in the Voigt profile analysis and serve only as a guide.} \\
     \multicolumn{12}{l}{$^a$The projected distances between different
             lensed sightlines at $z=0.7818$ are $(\overline{\rm A'C'},\overline{\rm A'D'}, \overline{\rm C'D'})=(7.8, 5.7, 5.7)$ kpc.} \\
     \multicolumn{12}{l}{$^b$Azimuthal angle of the QSO sightline from the major axis of the star-forming disk (measured counterclockwise).}\\
    \label{g2_table}
  \end{tabular}
\end{table*}
\end{center}

The top panel of Figure 7 displays the relative orientation of the
quadruply-lensed QSO to $G2$.  We have combined a high-contrast image
of the lensed QSOs with an image of $G2$ which is adjusted to
emphasize the spiral structures.  The relative spatial scale remains
the same as what is shown in Figure 1.  The green star symbols
indicate where the light from the QSO crosses the plane of $G2$.  We
designate the lensing corrected image positions as $A'$, $B'$, $C'$,
and $D'$.  In Table 3, we list for each sightline the lensing-modified
angular distance of $G2$ to the QSO image $\theta_{\rm qim}$ at the
$z=0.7818$, the corresponding projected distance $\rho$, the azimuthal
angle $\alpha$ of the QSO sightline with respect to the major axis of
the star-forming disk, the rest-frame absorption equivalent width
intergrated over all components $W_r(2796)$, the optical
depth-weighted mean velocity offset $\Delta\,v_\tau$, the velocity
width that encloses 90\% of the total line-of-sight effective optical
depth, $\delta\,v_{90}$, the number of components $n_c$ necessary to
reproduce the observed absorption profile along each sightline, and
the respective best-fit Voigt profile parameters of each component
[$\Delta\,v_c$, $\log\,N({\rm Mg\,II})$, $b_c$].

Combining known orientation of the QSO images to $G2$ with the
observed Mg\,II $\lambda\lambda$ 2796,2803 absorption profiles along
individual QSO sightlines leads a 3D velocity map of halo gas around
the galaxy.  The spectral images in the bottom of Figure 7 display the
effective optical depth $\tau_{\rm eff}$ of Mg\,II versus
line-of-sight velocity offset from $z=0.7818$ (the systemic redshift
of $G2$).  The values of $\tau_{\rm eff}$ are indicated by the color
bar in the lower-left corner.  We note that the projected distances
between different lensed sightlines at $z=0.7818$ are $(\overline{\rm
  A'C'},\overline{\rm A'D'}, \overline{\rm C'D'})=(7.8, 5.7, 5.7)$
kpc.  In constrast to the halo around $G1$, the absorbers uncovered
along sightlines at smaller projected separations from $G2$ show
distinct velocity structures with a difference in the mean velocity of
$\approx 90$ \kms.  Such distinct line-of-sight velocity distributions
suggests a turbulent halo gas near the star-forming disk of $G2$.

\begin{figure}
\begin{center}
\includegraphics[scale=0.45]{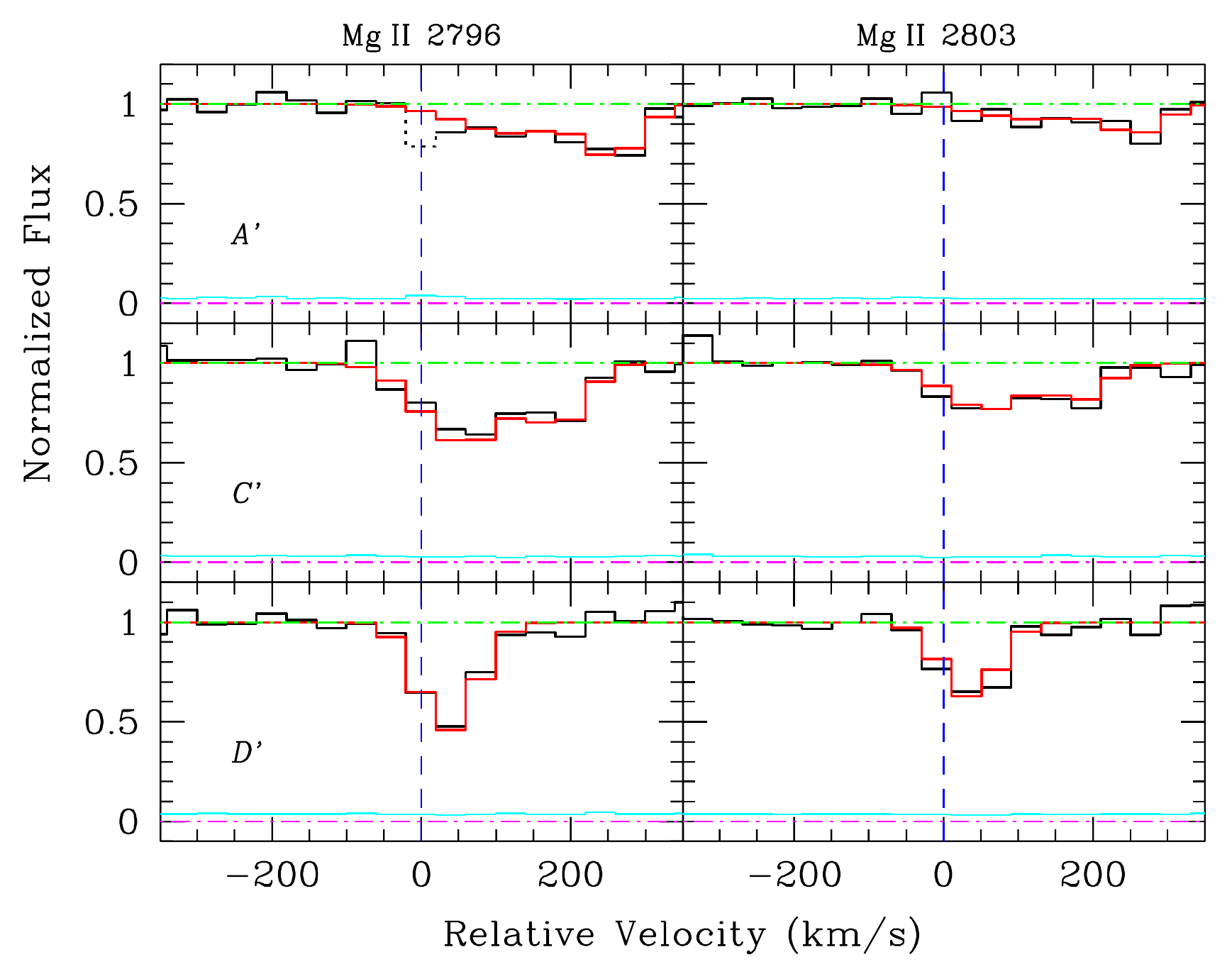}
\end{center}
\caption{Mg\,II $\lambda\lambda$ 2796,2803 absorption profiles of halo
  gas around the super-$L_*$ barred spiral galaxy $G2$ at $z=0.7818$.
  Absorption spectra of QSO sightlines $A'$, $C'$, and $D'$ (corrected
  lensed image positions at the redshift plane of $G2$; see the image
  in Figure 7) obtained with MagE are available.  Zero relative
  velocity corresponds to the systemic redshift of the galaxy at
  $z=0.7818$.  Despite a lower spectral resolution, the MagE spectra
  show remarkably distinct absorption morphology across different
  sightlines.  In particular, sightlines $C'$ and $D'$ occur at a
  similar projected distance from $G2$ and are roughly 6 kpc in
  projected separation.  The observed Mg\,II absorption is spread over
  130 \kms\ along the sightline toward $C'$, whereas more concentrated
  absorption is found within a narrow ${\rm FWHM}\approx 42$ \kms\
  along the sightline toward $D'$.}
\end{figure}

\begin{figure}
\begin{center}
\includegraphics[scale=0.47]{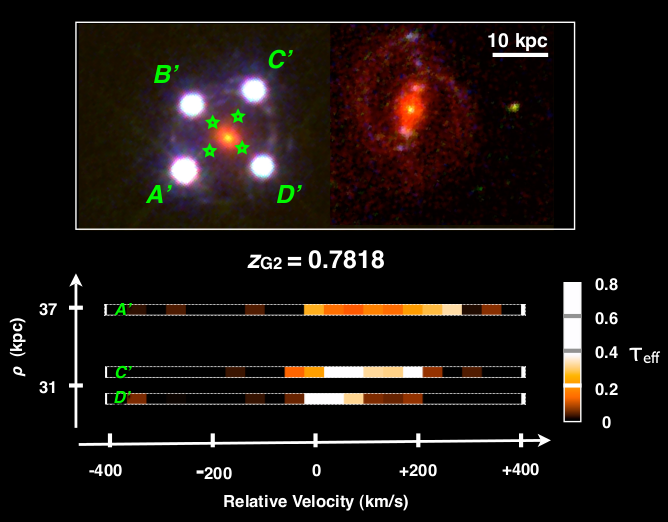}
\end{center}
\caption{Spatially resolved halo gas kinematics around the
  super-$L_*$, barred spiral galaxy $G2$ at $z=0.7818$.  The top panel
  shows the relative orientation of the quadruply-lensed QSO to $G2$
  at $\rho \sim 30$ kpc.  We have combined a high-contrast image of
  the lensed QSOs with an image of $G2$, adjusted to emphasize the
  spiral structures.  The relative spatial scale remains the same as
  what is shown in Figure 1.  In addition, because $G2$ is located
  behind the lens, the actual angular separations between the lensed
  QSO sightlines are smaller than what are marked by the lensed QSO
  images.  The green star symbols indicate where the light from the
  QSOs crosses the $z=0.7818$ plane.  Similar to Figure 5, the
  spectral images at the bottom display the effective optical depth
  $\tau_{\rm eff}$ of Mg\,II versus line-of-sight velocity offset from
  $z=0.7818$ (the systemic redshift of $G2$) for individual QSO
  sightlines (with increasing projected distances $\rho$ from bottom
  to top).  The values of $\tau_{\rm eff}$ are indicated by the color
  bar in the lower-left corner.  The projected distances between
  $\overline{\rm A'D'}$ and $\overline{\rm C'D'}$ are comparable,
  which are $\sim 6$ kpc.  But similar to what is found for the halo
  of $G1$, a larger velocity shear is seen with increasing $\rho$ than
  at comparable $\rho$.}
\end{figure}

\section{Discussion}

Combining known orientations of the quadruply-lensed QSOs to two
foreground galaxies with the observed Mg\,II $\lambda\lambda$
2796,2803 absorption profiles along individual QSO sightlines has
allowed us for the first time to resolve the kinematics of tenuous
halo gas on scales of $5-10$ kpc at $z>0.2$ (cf.\ Verheijen \etal\
2007).  We detect a Mg\,II absorber in every sightline through the
halo of each of the two galaxies in our study.  The absorber strength
observed in these galaxies is typical of what is seen in previous
close QSO--galaxy pair studies (e.g.\ Chen \etal\ 2010a,b).  While
both galaxies confirm previously reported high covering fraction of
Mg\,II absorbing gas at $\rho < 50$ kpc from star-forming disks, they
also present two contrasting examples of the complex halo gas
dynamics.

For $G1$, a likely member of a loose group as suggested by a
neighboring galaxy at a projected distance of 120 kpc away, a strong
Mg\,II absorber of $W_r(2796)>1$ \AA\ is present in all four
sightlines at $\approx 50$ kpc from the star-forming disk.  The
absorbers from different sightlines that are $\sim 10$ kpc apart in
projected separations share a similar asymmetry in their absorption
profiles.  We show a stacked halo velocity profile in the top panel of
Figure 8, which is established by coadding the observed effective
optical depth of Mg\,II along four different sightlines through the
halo of $G1$.  The coadded velocity profile is clearly inconsistent
with a Gaussian distribution, likely indicating that the absorbing gas
is not distributed randomly in the halo.  In addition, the velocity
width that encloses 90\% of the total line-of-sight effective optical
depth is $d\,v_{90} \approx 170$ \kms\ across all four sightlines with
a steady increase of $\Delta\,v_\tau$ from small ($B$ and $C$) to
large ($A$ and $D$) projected distances $\rho$.  Therefore, not only
does the absorbing gas appear to be deviating from a uniformly
distributed halo, but there also exists a strong spatial coherence
between these four sightlines.

In contrast, a moderately strong Mg\,II absorber of $W_r(2796)\approx
0.6$ \AA\ is detected in all three observed sightlines at $\approx 30$
kpc from $G2$, a barred spiral galaxy.  The Mg\,II absorbers uncovered
along individual sightlines that are $\sim 6$ kpc apart in projected
separations exhibit distinct absorption signatures, from a relatively
more concentrated single component found along the sightline toward
the $D'$ image with $\Delta\,v_\tau=+36$ \kms\ and
$\delta\,v_{90}=102$ \kms, to widely separated multiple components
with $\Delta\,v_\tau\apg 100$ \kms\ and $\delta\,v_{90}=280$ \kms.
While MagE delivers a spectral resolution of $\approx 70$ \kms\ and
therefore the intrinsic width of the single-component absorber
uncovered along the $D'$ sightline is likely to be $<102$ \kms, the
larger $\delta\,v_{90}$ seen along the sightlines toward $A'$ and $C'$
(which is four times the resolution element) can only be driven by the
underlying gas kinematics along these sightlines.  The vastly
different $\Delta\,v_\tau$ and $\delta\,v_{90}$ along these three
sightlines separated by $<10$ kpc indicate a turbulent velocity field
near $G2$.

The rich data set allows us to begin to constrain theoretical models
for inflows and galactic-scale outflows.  Here we examine different
models by comparing the observed 3D map of gaseous halos with model
expectations.  We also discuss the implications of our finding in the
general understanding of halo gas dynamics from absorption-line
observations.

\begin{figure}
\begin{center}
\includegraphics[scale=0.46]{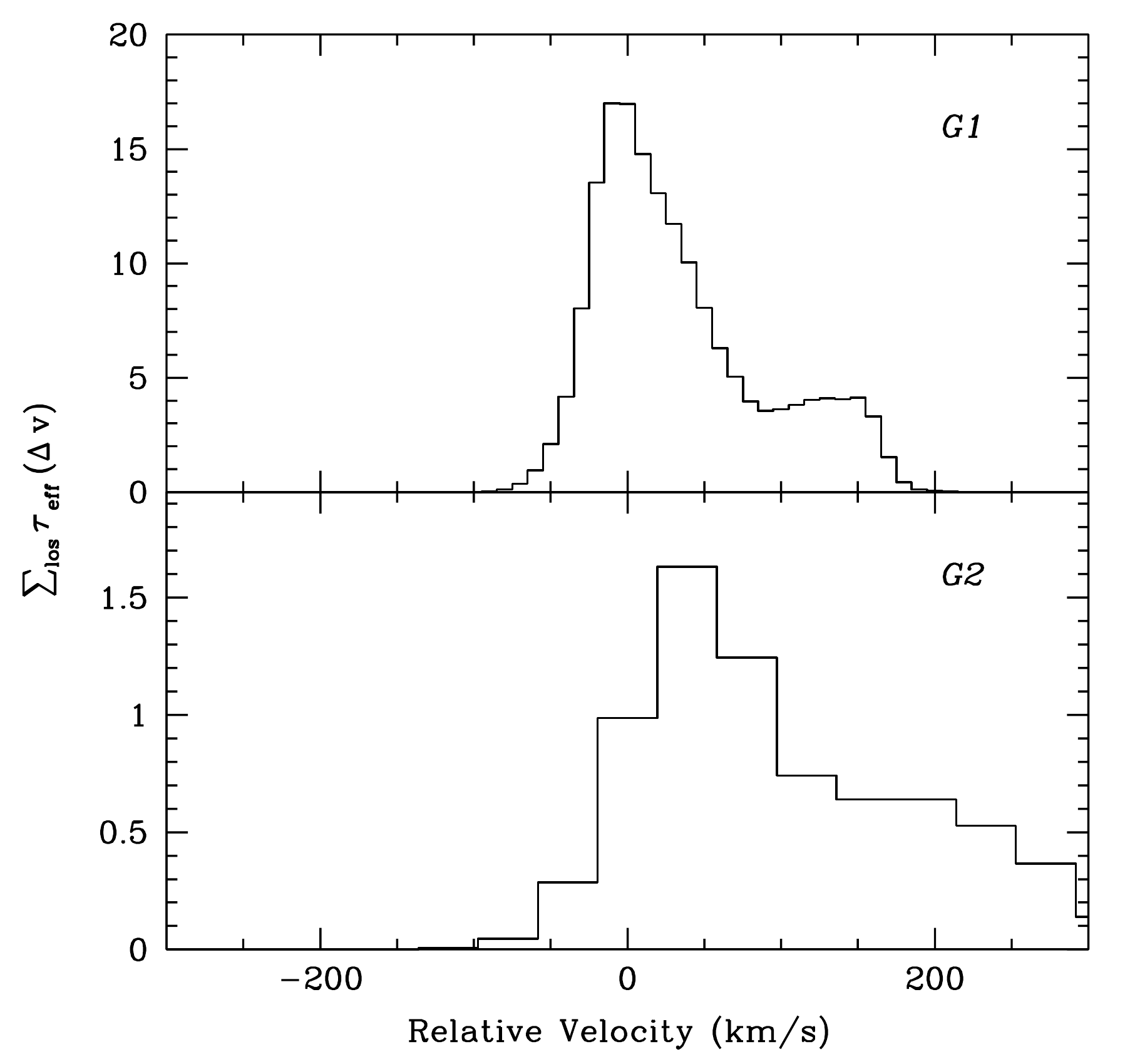}
\end{center}
\caption{Collective velocity profiles of the gaseous halos around $G1$
  (top) and $G2$ (bottom) from coadding the observed effective optical
  depth of Mg\,II versus velocity offset along different sightlines
  through each halo.  Same as Figures 4 and 6, the zero velocity in
  each panel corresponds to the systemic redshift of the associated
  galaxy.  The coadded velocity profiles from multiple sightlines are
  not a Gaussian function, indicating that the absorbing gas is not
  distributed randomly in each halo.}
\end{figure}

\subsection{Gas Flow Models}

We examine different gas flow models by comparing the observed 3D map
of gaseous halos with model expectations.  Three models are considered
here: (1) a rotating gaseous disk; (2) collimated outflows; and (3)
infalling streams or tidally stripped gas.

\subsubsection{A rotating gaseous disk}

We first consider the possibility that the gas probed by the Mg\,II
absorption follows the rotational motion of the star-forming disk
(e.g.\ Steidel \etal\ 2002; Chen \etal\ 2005).  Under this scenario,
the observed line-of-sight velocity offsets between different QSO
sightlines are interpreted as a result of velocity gradient along the
rotating disk.  Given the well-resolved optical morphologies of
galaxies $G1$ and $G2$, we are able to determine the disk inclination
($i_0$) and orientation ($\alpha$) relative to each QSO sightline and
deproject the observed line-of-sight velocity to the plane along the
stellar disk.  Following Chen \etal\ (2005), the projected distance
$\rho$ is related to galactocentric radius $R$ according to
\begin{equation}
\frac{R}{\rho}=\sqrt{1+\sin^2\alpha\,\tan^2i_0},
\end{equation}
and the line-of-sight velocity $v_{\rm los}$ is related to the
rotation speed $v_{\rm rot}$ along the disk according to (see Steidel
\etal\ 2002 for an alternative expression)
\begin{equation}
v_{\rm rot}=\frac{v_{\rm los}}{\cos\alpha\,\sin i_0}\sqrt{1+\sin^2\alpha\,\tan^2 i_0}.
\end{equation}
Equations (2) and (3) reduce to $R = \rho$ and $v_{\rm rot}=v_{\rm
  los}/\sin i_0$, when the sightline intercepts the major axis
($\alpha=0$).

For $G1$, sightlines $A$ and $B$ share a similar $\alpha$ and
therefore offer a direct measure of velocity gradient along the disk.
The observed projected distance between $A$ and $B$ of $\overline{\rm
  AB}=8.8$ kpc leads to a deprojected distance of $R_{AB}=10.9$ kpc
along the stellar disk.  The observed line-of-sight velocity
difference, $\Delta\,v_\tau(A)-\Delta\,v_\tau(B)=44\pm 14$ \kms, leads
to an increase in the rotation speeds of $dv_{\rm rot}^{AB}=169\pm 46$
\kms\ from $B$ to $A$.  Together, we derive a steep velocity gradient
of $\partial\,v/\partial\,R=16\pm 4$ \kms\ per kpc at $R>55$ kpc.
Similarly, sightlines $C$ and $D$ have the same $\alpha$.  With
$R_{CD}=11.8$ kpc and $dv_{\rm rot}^{CD}\approx 104$ \kms, we derive
$\partial\,v/\partial\,R\approx 9$ \kms\ per kpc at $R>50$ kpc.  The
velocity gradient would be even steeper, when considering the velocity
shear seen between $B$ and $C$ or between $A$ and $D$ that occur at
comparable galactocentric radii.  While large gaseous disks of radius
$>50$ kpc are known to exist, these galaxies also show a flat rotation
curve beyond the optical disks (e.g.\ Sofue \& Rubin 2001; Lelli
\etal\ 2010).

A simple rotating disk cannot explain the velocity map around $G2$,
either.  It is immediately clear that $C'$ and $D'$ occur at the same
projected distance to the star-forming disk and yet the absorbers
display a line-of-sight velocity difference of 60 \kms.  Considering
only component 1 along sightline $C'$ at $\Delta\,v_c=+57$ \kms\ would
still lead to a line-of-sight velocity difference of 22 \kms\ between
$C'$ and $D'$ at the same $\rho=31$ kpc.  The inferred velocity
gradient along the stellar disk between $D'$ and $A'$ would also be
enormous $\partial\,v/\partial\,R> 60$ \kms\ per kpc at $R>30$ kpc.
Therefore, we conclude that the gas kinematics revealed by the Mg\,II
absorption around either $G1$ or $G2$ is inconsistent with
expectations of a rotating disk.

\subsubsection{Bi-conical outflows}

Next, we consider the popular scenario that Mg\,II absorbers detected
in random QSO sightlines originate in supergalactic winds (e.g.\
Bouch\'e \etal\ 2006; Chelouche \& Bowen 2010; M\'enard \etal\ 2011;
Nestor \etal\ 2011).  In the local universe, large-scale galactic
outflows are commonly seen to follow a bi-conical pattern along the
rotation axis of the star-forming disk (e.g.\ Heckman et al.\ 2000)
with a varying degree of collimation, typically $2\,\theta_0=45^\circ -
100^\circ$ above the disk (Bland-Hawthorn, Veilleux \& Cecil 2007).
This is understood as the outflowing gas moves along the path of least
resistance.  In addition, observations of local starburst galaxies
have shown that galactic-scale superwinds exist in galaxies with a
global star formation rate per unit area exceeding $\sum_{\rm
  SFR}=0.1\ {\rm M}_\odot\,{\rm yr}^{-1}\ {\rm kpc}^{-2}$ (Heckman
2002).  Of the two galaxies identified in the foreground of
HE\,0435$-$1223, $G1$ has an unobscured star formation rate per unit
area of $\sum_{\rm SFR}(G1)\approx 0.06\ {\rm M}_\odot\,{\rm yr}^{-1}\
{\rm kpc}^{-2}$ after correcting for the disk inclination, and $G2$
has $\sum_{\rm SFR}(G2)\approx 0.002\ {\rm M}_\odot\,{\rm yr}^{-1}\
{\rm kpc}^{-2}$.  It is therefore possible that supergalatic winds are
present in at least $G1$, if dust extinction is significant.

Gauthier \& Chen (2012) developed an analytic formalism to fully
characterize the velocity gradient along biconical outflows based on
observed absorption profiles.  By attributing the observed
line-of-sight velocity spread to the intrinsic velocity gradient along
collimated outflows, these authors demonstrated that the outflow
velocity field can be uniquely established from absorption-line data,
when the outflow opening angle $\theta_0$, and the inclination ($i_0$)
and orientation ($\alpha$) angles of the star-forming disk are known.
They applied this analytic model to three edge-on disk galaxies that
have a background QSO probe the halo gas near the minor axis.  The
combination of highly inclined disks and a QSO ocurring near the minor
axis together helped tighten the constraints on the outflow velocity
field even with a single QSO sightline.  Gauthier \& Chen (2010)
showed that it is difficult for accelerated
outflows 
(e.g.\ Martin \& Bouch\'e 2009; Steidel \etal\ 2010; Murray \etal\
2011) to provide a general explanation for the strong Mg\,II absorbers
observed near the minor axis around the galaxies in their study.

Following Gauthier \& Chen (2012), we assume that the observed Mg\,II
absorbers in the quad-lens sightlines originated in collimated
outflows from the foreground galaxies $G1$ and $G2$, and derive the
velocity gradient necessary to explain the observed velocity shear
between different sightlines.  An important gain in our study here is
the available absorption profiles from multiple sightlines for each
galaxy.  An alternative model to collimated outflows is expanding
shells (e.g.\ Rauch \etal\ 2002) that resembles wind-blown bubbles
(e.g.\ Heiles 1979).  This alternative model is disfavored here due to
a lack of symmetry in the absorption profiles (cf.\ Bond \etal\ 2001)
between different sightlines about the systemic redshift of the
galaxies.

We first constrain $\theta_0$ (defined as half of the angular span of
the collimated outflows, see Figure 9) based on known inclination and
orientation angles of the disk and the observed relative velocity
offsets of different Mg\,II absorption components along different
lensed QSO sightlines.  We have shown that the star-forming disk of
$G1$ has $i_0=40^\circ$, $\alpha=120^\circ$ for sightlines $A$ and
$B$, and $\alpha=130^\circ$ for sightlines $C$ and $D$.  A strong
Mg\,II absorber is found in every one of the four sightlines.  With
the modest inclination of the disk and the sightlines intercepting the
halo at $\rho\approx 50$ kpc and $30^\circ-40^\circ$ from the minor
axis, we derive a minimum $\theta_0$ of $\theta_0^{\rm min}(G1)\approx
30^\circ$ in order for the collimated outflows to intercept all four
sightlines.  The maximum opening angle of the outflows for $G1$,
$\theta_0^{\rm max}(G1)$, is more uncertain, because of the complex
velocity distribution of individual components along different
sightlines.  Recall that no blueshifted components are detected along
the sightlines toward $A$ and $D$ and the dominant absorption along
sightlines $B$ and $C$ occurs near the systemic redshift of the
star-forming disk at $\rho\approx 45$ kpc (Figure 4 and Table 2).
Under a simple assumption that the outflowing gas is uniformly
distributed in the cone, the lack of blueshifted components toward $A$
and $D$ would lead to $\theta_0^{\rm max}(G1)\apl 55^\circ$, while the
presence of blueshifted components toward $B$ and $C$ would lead to
$\theta_0^{\rm max}(G1)\apg 55^\circ$ (Figure 9).  In particular,
because $A$ and $B$ occur at the same $\alpha$, it is expected that a
continuous stream moving outward and intercepting sightline $B$ to
produce the blueshifted component at $\Delta\,v_c=-23$ \kms\ would
continue to move on to intercept sightline $A$ at a similar
$\Delta\,v_c$ but the first absorbing component seen in sightline $A$
is nearly $+50$ \kms\ away.  The same caveat is seen $C$ and $D$ which
share a similar $\alpha=130^\circ$.  {\it If the blueshifted
  components seen in $B$ and $C$ arise in outflows that are moving
  toward the observer, then the outer layers of the collimated
  outflows are expected to be disrupted at $\rho \apl 50$ kpc} before
reach out to distances probed by $A$ and $D$.  However, some
outflowing gas would still have to move beyond 50 kpc in order to
produce the redshifted components see in $A$ and $D$.  {\it This would
  imply an increasing degree of collimation in the outflows with
  increasing distance.}  Alternatively, the blueshifted components
observed in sightlines $B$ and $C$ may not be associated with galactic
winds.  If attributing only redshifted components in all four
sightlines as due to outflows, then we can constrain the opening angle
with $\theta_0^{\rm max}(G1)\apl 55^\circ$.

For $G2$, we found that $i_0=25^\circ$ for the disk, and
$\alpha=26^\circ$ for sightlines $A'$ and $D'$, and $\alpha=16^\circ$
for sightline $C'$.  With a nearly face-on disk and the sightlines
intercepting the halo at $\rho\approx 30$ kpc and $\alpha\apl 26^\circ$ from
the major axis, we derive a minimum $\theta_0$ of $\theta_0^{\rm
  min}(G2)\approx 56^\circ$ in order for the collimated outflows to
intercept all three sightlines.
A large $\theta_0$ with $\theta_0 > 65^\circ$ would lead to
blueshifted components along sightlines $A'$ and $D'$ at $\rho \apg
30$ kpc, which are not seen.  Therefore, 
the maximum opening angle is constrained based on the absence of
blueshifted components at $\theta_0^{\rm max}(G2)\approx 65^\circ$.

\begin{figure}
\begin{center}
\includegraphics[scale=0.46]{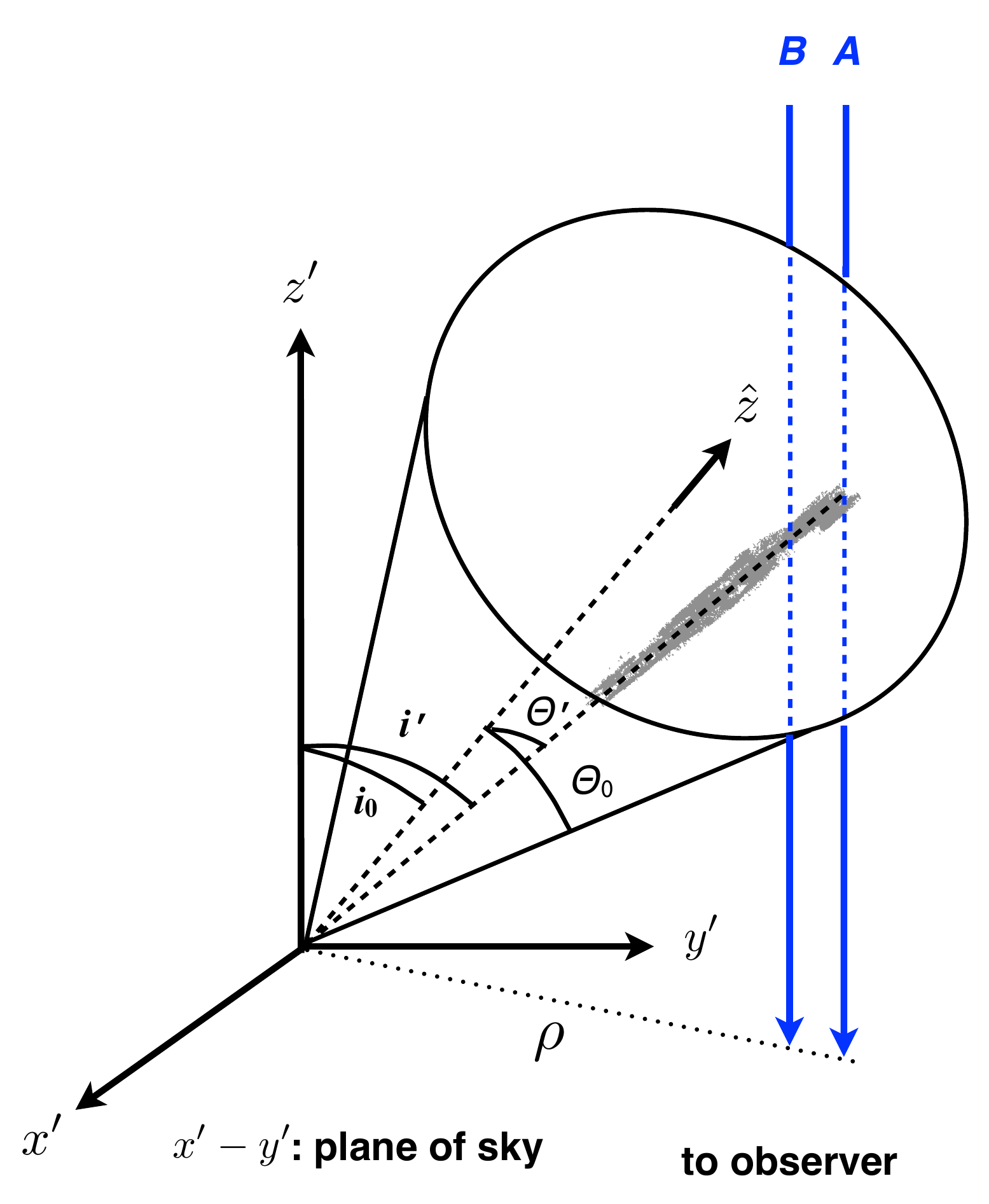}
\end{center}
\caption{Cartoon illustrating the impact geometry of multiple
  sightlines intercepting conical outflows.  The plane of sky is
  marked by the $x'-y'$ axes and the $z'$ axis points to distant objects.
  The outflow is reprepresentd by the cone oriented along the rotation
  axis $\hat{z}$ with inclination angle $i_0$ and opening angle
  $2\times\theta_0$.  It is expected from this orientation that when
  $\theta_0$ is sufficiently large to cross the plane of the sky, the
  outflows would imprint blueshifted absorption features in the
  spectra of background sources.  A lack of blueshifted components
  would therefore constrain the maximum $\theta_0$.  Within the cone,
  gas streams move outward at different angle $\theta'\apl \theta_0$
  from $\hat{z}$ with corresponding inclination angle $i'$ (the thick
  paint brush line is marked as an example of such outward moving
  stream).  In principle, the velocity gradient along the streams can
  be directly constrained by the average velocity shears observed
  between different sightlines (marked $A$ and $B$) that occur at the
  same azimuthal angle on the sky (Equation 4).}
\end{figure}

Next, we calculate the velocity gradient in the collimated outflows
that is necessary to produce the observed velocity shear between
different sightlines.  Because the relative locations of individual
components between different sightlines in the collimated outflows are
not known, we compute the velocity gradient based on $\Delta\,v_\tau$,
the effective optical depth weighted line center, average over the
entire absorber in each sightline.  As mentioned earlier, sightlines
$A$ and $B$ intercept the halo of $G1$ at a nearly identical azimuthal
angle $\alpha=120^\circ$ from the disk and are roughly 9 kpc apart.
It is therefore conceivable that $A$ and $B$ trace the same streams of
gas outflowing from the disk at some mean angle $\theta'$ from the
disk rotation axis (see Figure 9 for illustrations).  But recall the
caveat that the blueshifted components uncovered toward $B$ and $C$
may not be associated with outflows, in which case the velocity
gradient inferred below would represent an upper limit.

Adopting the analytic conical outflow model of Gauthier \& Chen
(2012), we convert the line-of-sight velocity gradient to outflow
velocity along the rotation axis
following
\begin{equation}
\frac{\delta\,v}{\delta\,z} = \frac{\cos i'\,\sin i'}{\cos\theta'}\,\frac{\delta\,v_{\rm los}}{\delta\,\rho},
\end{equation}
where $\delta\,v_{\rm los}=\Delta\,v_\tau(A) - \Delta\,v_\tau(B)$,
$\delta\,\rho=\rho(A) - \rho(B)$, $\theta'$ is the angle of the
outflowing stream away from the rotation axis and is related to $\rho$
and $\alpha$ according to
\begin{equation}
\tan\theta'=\frac{\sin i_0\cos\alpha}{\sin\alpha},
\end{equation}
and $i'$ is the corresponding inclination angle of the stream at
$\theta'$ and is related to $\rho$ and $\theta'$ according to
\begin{equation}
\sin i'=\frac{\rho}{z}\cos\theta'=\frac{\sin i_0\cos\theta'}{\sin\alpha}.
\end{equation}
Note that $\theta'=0$ and $i'=i_0$, if $\alpha=90^\circ$ (namely when
the sightlines occur near the minor axis of the disk).

Comparing $\Delta\,v_\tau$ along $A$ and $B$ sightlines leads to a
velocity difference of $\delta\,v=+26$ \kms\ from $60$ kpc to $70$ kpc
in $z$-height for an outflowing stream at $\theta'\approx 20^\circ$
from the rotation axis.  Likewise, sightlines $C$ and $D$ intercept
the halo of $G1$ at $\alpha=130^\circ$ and are 10 kpc apart.  Equation
(4) leads to to a velocity difference of $\delta\,v=+20$ \kms\ from 50
kpc to 60 kpc in $z$-height for an outflowing stream at
$\theta'\approx 28^\circ$ from the rotation axis.  The relatively
shallow velocity gradient is comparable to the expected acceleration
in radiation pressure driven winds at large distances (e.g.\ Murray
\etal\ 2011).  At the same time, the deprojected outflow speed ($\sim
100$ \kms) at $>50$ kpc above the disk is significantly smaller than
expectations from accelerated outflows (e.g.\ Steidel \etal\ 2010;
Murray \etal\ 2011), but comparable to expectations for galactic winds
driven by cosmic rays (e.g.\ Booth \etal\ 2013).

For $G2$, sightlines $A'$ and $D'$ intercept the halo at a nearly
identical azimuthal angle $\alpha=26^\circ$ from the disk and are
roughly 6 kpc apart.  Considering the Mg\,II absorbers uncovered in
these two sightlines, we calculate the velocity gradient in the
collimated outflows based on the observed $\Delta\,v_\tau$ and obtain
a steep velocity gradient $\delta\,v/\delta\,z=18\,\kms\,/\,{\rm kpc}$
at $30-40$ kpc above the disk.  The inferred velocity gradient exceeds
expectations for accelerated outflows at these distances (e.g.\
Steidel \etal\ 2010; Murray \etal\ 2011).

In summary, our analysis shows that a simple outflow scenario alone
cannot provide a consistent explanation for {\it all} absorbing
components found in either $G1$ or $G2$.  This is qualitatively
consistent with the low metallicity ($0.1-0.4$ solar for $G1$ and
$0.06$ solar for $G2$) inferred for the gas based on the observed
$W_r(2796)$.  Under the assumption that the Mg\,II absorbers uncovered
in the vicinities of these galaxies originate in collimated outflows,
we have demonstrated that multiple probes afforded by the quad-lens
allow us to constrain both the outflow opening angle and the velocity
gradient.  The constraints are based entirely on the relative
locations between the lensed QSOs and the line-of-sight absorption
profiles, without further assumptions for how individual components
along different sightlines are associated with one another.

For $G1$, attributing all the observed Mg\,II components to collimated
outflows would require an increased degree of collimation with
increasing distance in the outflows.  If only the redshifted
components in all four sightlines are attributed to collimated
outflows, then the outflow opening angle would be constrained by the
data to be within $\theta_0=[30^\circ,55^\circ]$.  In turn, a shallow
velocity gradient of $\delta\,v/\delta\,r\apl 2\,\kms\,/\,{\rm kpc}$
at $\apg 50$ kpc above the star-forming disk is derived which is
comparable to model expectations for supergalactic winds, while the
deprojected velocity falls well below the model expectations.  Such
discrepancy between outflowing speed and the velocity gradient
suggests deceleration rather than acceleration in the outflows.  

For $G2$, attributing all the observed Mg\,II components to collimated
outflows constrains the outflow opening angle to within
$\theta_0=[56^\circ,65^\circ]$.
The inferred velocity gradient would substantially exceed model
expectations, requiring additional energy input at $\apg 30$ kpc above
the disk.  Taking into account the observed low $\sum_{\rm SFR}$ in
$G2$, we therefore find it unlikely that the a large fraction of
Mg\,II absorbers observed in all three sightlines at $>30$ kpc from
$G2$ originate in supergalactic winds.

\subsubsection{Infalling streams or stripped gas}

We have shown that neither rotating disks, nor collimated outflows can
provide a single, consistent model for explaining the observed Mg\,II
absorption kinematics along multiple sightlines around $G1$ and $G2$.
At the same time, the absorption profiles exhibit a strong spatial
coherence on scales of $\sim 10$ kpc, suggesting the presence of bulk
flows in the halos.  The bulk flows are characterized for $G1$ by (1)
a similar asymmetric absorption profile of comparable velocity width
($\delta\,v_{90}\approx 170$ \kms) across four different sightlines
(Figure 8), (2) a systematic velocity increment of $\delta\,v\approx
+40$ \kms\ observed from small ($B$ and $C$ at $\rho=44$ kpc) to large
($A$ and $D$ at $\rho=54$ kpc) projected distances, as well as (3) a
smaller velocity difference (by a factor of $\approx 2$) observed
between sightlines of similar projected distances and $\approx 10$ kpc
apart ($\delta\,v_{\rm los}({\rm BC})=14$ \kms\ at $\rho \approx 44$
kpc and $\delta\,v_{\rm los}({\rm AD})=22$ \kms\ at $\rho\approx 54$
kpc).  Similarly for $G2$, a large velocity increment of
$\delta\,v\approx +150$ \kms\ is observed from small ($D'$ at
$\rho=31$ kpc) to large ($A'$ at $\rho=37$ kpc) projected distances,
and the velocity difference observed between $C'$ and $D'$ at
$\rho=31$ kpc and $\approx 6$ kpc apart is a factor of three smaller,
with $\delta\,v_{\rm los}({\rm C'D'})=60$ \kms.

Here we consider a third scenario in which Mg\,II absorbing gas
originates in gaseous streams driven by gravitational forces in the
halo.  Gaseous streams can be present either due to accretion from the
intergalactic medium (IGM) or tidal/ram-pressure stripping from
interacting galaxies.  We first consider streams accreted from the
IGM.

Galaxies are commonly believed to grow in mass by accretion and
mergers (e.g.\ White \& Rees 1978).  But while renewed emphases (e.g.\
Kere$\check{\rm s}$ \etal\ 2005, 2009; Birnboim \& Dekel 2003; Dekel
\& Birnboim 2006; Dekel \etal\ 2009; Fumagalli \etal\ 2011) have been
made on the importance of feeding galaxy growth by cold streams from
the IGM, few direct detections of accretion have been reported in
distant galaxies (e.g.\ Rauch \etal\ 2011; Rubin \etal\ 2012).  Though
indirect evidence suggesting co-rotational motion between gas and
stellar disks has been reported by various authors (e.g.\ Steidel
\etal\ 2002; Chen \etal\ 2005; Kacprzak \etal\ 2011; Bouch\'e \etal\
2013).  Although QSO absorbers uncovered in the vicinities of galaxies
in principle offer a promising candidate for infalling gas clouds,
ambiguities arise when starburst driven outflows are thought to
contribute a dominant fraction of these QSO absorbers.

The lack of direct detections of gas accretion has been attributed to
a low covering factor ($\sim 10$\%) of cold streams around galaxies
(e.g.\ Faucher-Gigu\`ere \& Kere$\check{\rm s}$ 2011; Fumagalli \etal\
2011).  In contrast, we have uncovered a Mg\,II absorber in every
observed sightline through each of the two galaxies studied here.  Our
multi-sightline observations have revealed not only that extended halo
gas is clearly present, but also that the rate of incidence in Mg\,II
absorption is qualitatively consistent with being 100\% (e.g.\ Chen
\etal\ 2010a).  While the observed 100\% covering fraction appears to
be discrepant from model expectations for cold accretion, a direct
comparison with simulations is difficult due to unknown neutral
hydrogen column densities of the gas revealed through Mg\,II
absorption.

On the other hand, the line-of-sight velocity widths
($\delta\,v_{90}$) of the observed Mg\,II absorbers are roughly
consistent with the circular velocity of the halo, which is related to
halo mass according to $v_{\rm circ}\sim
144\,M_{12}^{1/3}\,(1+z)^{1/2}$ \kms\ with $M_{12}$ representing halo
mass in units of $10^{12}\,{\rm M}_\odot$.  $G1$ is estimated to
reside in a halo of $10^{12}\,{\rm M}_\odot$, which at $z=0.4188$ has
$v_{\rm circ}\approx 170$ \kms.  The Mg\,II absorption profiles along
four sightlines near $G1$ show a velocity width of
$\delta\,v_{90}\approx 160-180$ \kms.  $G2$ is estimated to reside in
a halo of $5\times 10^{12}\,{\rm M_\odot}$, which at $z=0.7818$ has
$v_{\rm circ}\approx 330$ \kms.  The Mg\,II absorption profiles along
three sightlines near $G2$ show a velocity width of
$\delta\,v_{90}\approx 100-280$ \kms.  Therefore, gravitation motion
of infalling streams through the halo can explain the observed
velocity widths in Mg\,II absorption.

Alternatively, gaseous streams can be present due to tidal
interactions (e.g.\ Yun \etal\ 1994; Putman \etal\ 1998) or
ram-pressure stripping (e.g.\ Lin \& Faber 1983).  In particular, $G1$
has a known neighboring galaxy at a projected distance of 120 kpc away
(\S\ 3.1), and available HST images show features resembling tidal
debris at $\approx 3''$ south and $\approx 2''$ northwest of $G1$
(Figure 1; Morgan \etal\ 2005).  Detailed H\,I maps of nearby galaxy
groups from 21~cm observations (e.g.\ Yun \etal\ 1994; Hunter \etal\
1998; Chynoweth \etal\ 2008; Mihos \etal\ 2012) have revealed extended
H\,I gas covering a large fraction of the area within radius $\approx
50$ kpc from the star-forming disk, and the relative velocities of the
tidal arms are found to span a range of $\approx \pm\,200$ \kms\ from
the systemic redshift of the galay group.  Given the dynamic range of
gas motion in stripped gas, we expect to find a mean velocity spread
of $\sim 230$ \kms\ projected along a random sightline which is
consistent with what is measured for the Mg\,II absorbers.

In summary, the modest velocity widths $\delta\,v$ of Mg\,II absorbers
observed along multiple sightlines are consistent with the
expectations of absorption produced either by infalling streams from
the IGM or by stripped gas from tidal interaction or ram pressure.  We
note that a modest velocity width is also expected for absorption due
to recycled winds (e.g.\ Oppenheimer \etal\ 2010), but {\it it is not
  clear whether recycled winds can explain the asymmetric absorption
  profiles observed both along individual sightlines and in coadded
  spectra} (Figure 8).  It appears that the gas kinematics revealed
along multiple sightlines around $G1$ and $G2$ are best described by
gaseous streams of $d_s\apg 10$ kpc in width.

\subsection{Implications}

We have considered three different scenarios, (1) a rotating gaseous
disk, (2) collimated outflows, and (3) infalling streams or stripped
gas, as possible explanations for the observed Mg\,II absorbers along
multiple sightlines around $G1$ and $G2$.  We find that attributing
the observed line-of-sight velocity differences across multiple
sightlines to a rotating disk would imply an unreasonably steep
velocity gradient at $>50$ kpc.  A rotating disk model is therefore
ruled out.  At the same time, collimated outflows cannot fully explain
the observed velocity distribution across multiple sightlines.  While
it is conceivable that some fraction of the observed absorption
components originate in collimated outflows with the rest originating
in gas infall, the inferred energetics is significantly lower than
expectations for supergalactic winds driven by young stars (but
comparable to cosmic-ray driven winds).  In contrast, gaseus streams
of $d_s\apg 10$ kpc in width due to either accretion from the IGM or
tidal/ram-pressure stripping offer the best model to fully explain the
gas kinematics revealed by the observed Mg\,II absorption profiles.
We also note that the inferred metal enrichment level (between 0.06
and 0.4 solar) of the Mg\,II gas from $W_r(2796)$ is also
significantly smaller than super-solar metallicity expected for the
interstellar medium of massive galaxies with $M_*>2\times
10^{10}\,{\rm M}_\odot$ (e.g.\ Savaglio \etal\ 2005), but consistent
with expectations for infalling streams or stripped gas from
satellites or the outskirts of a star-forming disk.  

Here we discuss the implications of our observations for infall
models.  Specifically, we focus on the turbulence and mass flow rate
in infalling streams.

\subsubsection{Turbulence in gaseous halos}

While gaseous streams from either accretion or tidal debris provide a
promising explanation for the observed velocity widths and relative
motion between different sightlines, a clear discrepancy between the
observations and model expectations is the gas covering fraction.
Specifically, the covering fraction of infalling material is expected
to be low, $\sim 10$\%, through out galactic halos (e.g.\
Faucher-Gigu\`ere \& Kere$\check{\rm s}$ 2011; Fumagalli \etal\ 2011).
This is in contrast to the occurrence of a relatively strong Mg\,II
absorber in every observed sightline through the two galaxies in our
study.

A potential explanation for the discrepancy between observations and
simulation predictions is turbulence.  Whether or not the streams are
turbulent is assessed by the Reynolds number ($Re$) of the flows.  For
photo-ionized gas at temperature $T \sim 10^4$ K that is expected for
infalling streams (e.g.\ van de Voort \& Schaye 2012; Rosdahl \&
Blaizot 2012), $Re$ is expected to be very large at $Re\apg 10^9$
(Andrey Kravtsov, private communication).  In contrast to the laminar
flows commonly shown in numerical simulations, the cold streams should
be turbulent.  Consequently, some fraction of the kinetic energy is
expected to be converted into internal energy resulting in a more
chaotic state of the flows, particularly in lower density part of the
streams.

Empirically, the amount of turbulence in a gas can be determined from
comparing the observed absorption line widths of two different atoms,
because the line width is characterized by the Doppler parameter $b$
which combines the effect of thermal broadening and bulk motion.  Here
we cannot use this approach to constrain the turbulent motion with
only Mg\,II lines available and no additional information on the gas
temperature.  But in principle, we can estimate the degree of
turbulent motion based on the observed velocity dispersion between
different absorption components along different sightlines (e.g.\
Rauch \etal\ 2002).

On the other hand, the velocity dispersion observed along individual
sightlines should indicate a combined effect of small-scale ($\sim 10$
kpc) turbulent motion and large-scale ($\sim 100$ kpc) gravitational
acceleration.  The comparable scale between the expected $v_{\rm
  circ}$ and observed $\delta\,v_{90}$ has led us to attribute the
observed velocity dispersion along individual sightlines largely to
gravitational acceleration, which implies a significantly smaller
effect of turbulent motion on the observed line-of-sight velocity
dispersion.

To obtain a crude estimate for the amount of turbulence in the
infalling streams, we instead compare the difference in the optical
depth weighted mean velocity ($\Delta\,v_\tau$) between pairs of
sightlines.  We restrict the comparison to only pairs that occur at
similar projected distances, because of the apparent coherence in the
observed velocity gradient with increasing distance (e.g.\ from $B$
and $C$ at $\rho\approx 44$ kpc to $A$ and $D$ at $\rho\approx 54$ kpc
around $G1$).  

For $G1$, sightlines $B$ and $C$ at $\rho\approx 44$ kpc are separated
by $d_{\rm BC}\approx 8$ kpc in projected distance and the difference
in $\Delta\,v_\tau$ is $|\Delta\,v_\tau(B) - \Delta\,v_\tau(C)|=14$
\kms.  Similarly, sightlines $A$ and $D$ at $\rho\approx 54$ kpc are
separated by $d_{\rm BC}\approx 10$ kpc and the difference in
$\Delta\,v_\tau$ is $|\Delta\,v_\tau(A) - \Delta\,v_\tau(D)|=22$ \kms.
Together, we estimate the velocity dispersion on scales of $d_s\sim
10$ kpc as a result of turbulent motion is of order $v_{\rm
  disp}\sim\sqrt{3}\,\Delta\,v_\tau\approx 35$ \kms\ at $\rho\approx
50$ kpc (roughly $1/3$ of the virial radius) from $G1$.

For $G2$, sightlines $C'$ and $D'$ occur at $\rho=31$ kpc and are
separated by $d_{\rm C'D'}\approx 6$ kpc in projected distance.  The
Mg\,II absorbers uncovered along the two sightlines exhibit distinct
kinematic features with $\delta\,v_{90}=280$ and 102 \kms for $C'$ and
$D'$, respectively.  The distinct absorption profiles on scales of
$\sim 6$ kpc already signal the presence of strong turbulence.  Even
if considering only the blue component in $C'$ (component 1 in Table
3) as originating in the same stream, the velocity difference would be
$|\Delta\,v_{c}^1(C') - \Delta\,v_{c}^1(D')|=22$ \kms\ on scales of
$d_s\sim 6$ kpc at $\rho=31$ kpc (roughly $1/6$ of the virial radius)
from $G2$.

In summary, the observed velocity offsets between different sightlines
suggest that halo gas around $G1$ and $G2$ is turbulent.
The amount of turbulence is characterized by a velocity dispersion of
$\sim 35$ \kms\ on scales of $6-10$ kpc at $\approx 30-50$ kpc from
the star-forming disk.

\subsubsection{Mass flow rate}

A particularly interesting quantity regarding gaseous streams in
galactic halos is the mass flow rate, which can be compared with model
predictions.  {\it However, there are a number of caveats that make an
accurate estimate of this quantity difficult.  For example, the size
and orientation of the streams are not known.  In addition, the
streams in our study here are revealed by the presence of Mg\,II
absorption doublet.  Both the ionization fraction and metallicity of
the gas are required in order to infer the total mass contained in the
Mg\,II absorbers.}  As described in \S\,4.1, we assume an ionization
fraction of $f_{\rm Mg^+}=0.1$ based on a simple ionization model
presented in Chen \& Tinker (2008) and infer a chemical enrichment
level of $f_{Z}=0.1-0.4$ solar metallicity for the Mg\,II absorbers
around $G1$ and $f_{z}\approx 0.06$ solar for the Mg\,II absorbers
around $G2$ based on the empirical metallicity--$W_r(2796)$ relation
of Murphy \etal\ (2007).

The mass flow rate of streams probed by Mg\,II absorbers is
characterized by the product of Mg\,II column density and
cross section of the streams following,
\begin{eqnarray}
\dot{M}&\equiv&\frac{\Delta\,M}{\Delta\,t}
=\frac{1}{f_{\rm Mg^+}}\frac{1}{f_Z}\left(\frac{\rm Mg}{\rm H}\right)_\odot^{-1}\frac{m_{\rm Mg}\,N({\rm Mg\,II})\,A_s\,(\cos\,i')\,v_s}{d\,\ell} \nonumber \\
&=&\frac{1}{f_{\rm Mg^+}}\frac{1}{f_Z}\left(\frac{\rm Mg}{\rm H}\right)_\odot^{-1}\frac{m_{\rm Mg}\,N({\rm Mg\,II})\,A_s\,v_{\rm los}}{d\,\ell},
\end{eqnarray}
where $m_{\rm Mg}$ is the atomic weight of magnesium,
$A_s\equiv\pi\,d_s^2/4$ is the cross section of the streams, $i'$ is
the orientation of infalling streams relative to the line of sight,
$v_s$ is the velocity of the streams, and $d\ell$ is the
characteristic length of the streams probed by the sightline.
Adopting a characteristic size of streams $d\ell\sim d_s$, we can
approximate the mass flow rate as
\begin{eqnarray}
\dot{M}
&\sim&\frac{\pi}{4}\frac{1}{f_{\rm Mg^+}}\frac{1}{f_Z}\left(\frac{\rm Mg}{\rm H}\right)^{-1}\,m_{\rm Mg}\,N({\rm Mg\,II})\,d_s\,v_{\rm los} \\
&\approx& 4\times \frac{0.1}{f_{\rm Mg^+}}\frac{0.1}{f_Z}\frac{N({\rm Mg\,II})}{10^{13}\cmjj}\frac{d_s}{10\,{\rm kpc}}\frac{v_{\rm los}}{100\,{\rm km}/{\rm s}}\,\frac{{\rm M}_\odot}{{\rm yr}}.
\end{eqnarray}
For an observed line-of-sight velocity of $v_{\rm los}\approx 50$
\kms\ and a characteristic dimension of $d_s=10$ kpc for the infalling
streams probed by the Mg\,II absorbers, the inferred mass flow rate is
$\dot{M}\sim 2\,{\rm M}_\odot\,{\rm yr}^{-1}$.  

\section{Summary}

In conclusion, we have established spatially resolved velocity maps on
scales of $5-10$ kpc for two galaxies at $z=0.4-0.8$ using absorption
spectroscopy of quadruply-lensed QSO HE\,0435$-$1223 at $\rho\apl 50$
kpc from the galaxies.  The analysis presented here demonstrates that
multiple-QSO probes enable studies of spatially resolved gas
kinematics around distant galaxies, which provide key insights into
the physical nature of circumgalactic gas beyond the nearby universe.
The main results of our study are summarized as the following:

(1) The first galaxy $G1$ at $z=0.4188$ and $\rho\approx 50$ kpc from
the quad-lens is best characterized as a typical, star-forming $L_*$
galaxy with an on-going star formation rate of ${\rm SFR}\sim 4\,{\rm
  M}_\odot\,{\rm yr}^{-1}$ and total stellar mass of $M_*\approx
(2-3)\times 10^{10}\,{\rm M}_\odot$.  A strong Mg\,II absorber of
$W_r(2796)>1$ \AA\ is detected in everyone of the four lensed QSO
sightlines in the vicinity of the galaxy, indicating a high gas
covering fraction and suggesting a chemical enricment level of
$0.1-0.4$ solar in the halo gas.  The Mg\,II absorption doublet is
generallly characterized by a dominant component near the systemic
velocity, which is followed by secondary absorbing components at $\sim
100$ \kms\ in the red.  Such asymmetric kinematic signatures apply to
all four sightlines separated by $8-10$ kpc in projected distances,
and only a relatively small velocity shear (between $\Delta\,v\approx
20$ \kms) is seen across these different sightlines.  The absorption
profiles exhibit a strong spatial coherence on scales of $\sim 10$
kpc, suggesting the presence of bulk flows in the halos.

(2) The second galaxy $G2$ at $z=0.7818$ and $\rho\approx 30$ kpc from
the quad-lens is best characterized as a massive, super-$L_*$ galaxy
that resembles quiescent star-forming galaxies at $z=0.5-1$.  A
moderately strong Mg\,II absorber of $W_r(2796)=0.5-0.7$ \AA\ is
detected in everyone of the three lensed QSO sightlines observed in
the vicinity of the galaxy, also indicating a high gas covering
fraction and suggesting a chemical enricment level of $\approx 0.06$
solar in the halo gas.  While the Mg\,II absorption doublet around
$G2$ also exhibit an asymmetric profile, distinct kinematic signatures
are observed between different sightlines separated by $\sim 6$ kpc,
suggesting a more turbulent nature of the halo gas.

(3) Interpreting the observed velocity shear around either $G1$ or
$G2$ as a result of an underlying rotating gaseous disk leads to a
velocity gradient as steep as $\partial\,v/\partial\,R=16\pm 4$ \kms\
per kpc at $R>55$ kpc, which is not seen in any nearby galaxies.  We
therefore conclude that the gas kinematics revealed by the MgII
absorption around either G1 or G2 is inconsistent with expectations of
a rotating disk.

(4) Interpreting the observed line-of-sight velocity shear across all
four sightlines near $G1$ as a result collimated outflows would imply
an increasing degree of collimation in the outflows with increasing
distance.  In addition, the inferred outflow speed is comparable to
expectations from cosmic-ray driven winds, but is significantly
smaller than expectations from accelerated outflows.  While a single
collimated outflows model cannot fully explain the spatially resolved
gas kinematics around $G1$, it is conceivable that some fraction of
the observed absorption components originate in collimated outflows.
However, the inferred metallicity of the Mg\,II gas appears too low,
in comparison to a solar metallicity (or higher) expected for the ISM
gas of massive galaxies.  In contrast, the inferred velocity gradient
for $G2$ would substantially exceed model expectations. It is
therefore unlikely that collimated outflows can explain the spatially
resolved gas kinematics around $G2$.

(5) The strong spatial coherence in Mg\,II absorption across multiple
sightlines is best explained by gaseous streams of $\apg 10$ kpc in
width driven by gravitational forces in the halo, either due to
accretion from the IGM or stripped gas from interacting galaxies.
This is supported for $G1$ by the presence of a nearby companion at
120 kpc away and by the low metal content inferred from the absorber
strength.  The absorption kinematics between sightlines at similar
projected distances to the star-forming disk exhibit a velocity offset
of $\approx 20$ \kms.  Interpreting the velocity difference as a
result of turbulent motion leads to an estimate for the amount of
turbulence in halo gas of $\sim 35$ \kms\ in velocity dispersion on
scales of $6-10$ kpc at $\approx 30-50$ kpc from the star-forming
disk.


\section*{Acknowledgments}

It is a pleasure to thank Denis Erkal, Nick Gnedin, Andrey Kravtsov,
Lynn Matthews, and Michael Rauch for helpful discussions.  We thank
Chris Kochanek for providing the best-fit parameters of their lens
model.  We also thank the staff of the Las Campanas Observatory for
their expert assistance with the observations.  JRG gratefully
acknowledges the financial support of a Millikan Fellowship provided
by Caltech.  KS acknowledges support from the University of Michigan's
President's Postdoctoral Fellowship.


\label{lastpage}

\end{document}